\documentclass[12pt]{JHEP3}
\usepackage{amssymb} 
\usepackage{amsmath}
\usepackage{mathtools}
\usepackage{amsfonts}    
\usepackage{dsfont}
\usepackage{young}
\usepackage[vcentermath]{youngtab}

\newcommand{\be}{ \begin{equation}}
\newcommand{\ee}{\end{equation}}
\newcommand{\bea}[1]{\begin{eqnarray}\label{#1} }
\newcommand{\eea}{\end{eqnarray}}

\def\ZZZ{{\hskip-3pt\hbox{ Z\kern-1.6mm Z}}}

\def\zzz{{\hskip-3pt\hbox{ z\kern-1mm z}}}

\def\one{{\hbox{ 1\kern-.8mm l}}}
\def\zero{{\hbox{ 0\kern-1.5mm 0}}}

\title{The Worldsheet Dual of Free Super Yang-Mills in 4D}

\author{
Matthias R.\ Gaberdiel$^{a}$ and Rajesh Gopakumar$^{b}$ \\ 
$^a$Institut f\"ur Theoretische Physik, ETH Zurich, \\
$\;$CH-8093 Z\"urich, Switzerland \\
$\;$\email{gaberdiel@itp.phys.ethz.ch}\\ \\ 
$^b$  International Centre for Theoretical Sciences (ICTS-TIFR),\\
$\;$Shivakote, Hesaraghatta Hobli, \\
$\;$Bangalore North, India 560 089\\ 
$\;$\email{rajesh.gopakumar@icts.res.in}}

\abstract{The worldsheet string theory dual to free 4d ${\cal N}=4$ super Yang-Mills theory was recently proposed in \cite{Gaberdiel:2021iil}. It is described by a free field sigma model on the twistor space of ${\rm AdS}_5\times {\rm S}^5$, and is a direct generalisation of the corresponding model for  tensionless string theory on ${\rm AdS}_3\times {\rm S}^3$. As in the case of ${\rm AdS}_3$, the worldsheet theory contains spectrally flowed representations. We proposed in \cite{Gaberdiel:2021iil} that in each such sector only a finite set of generalised zero modes (`wedge modes') are physical.  Here we show that after imposing the appropriate residual gauge conditions, this worldsheet description reproduces precisely the spectrum of the planar gauge theory. More specifically, the states in the sector with $w$ units of spectral flow match with single trace operators built out of $w$ super Yang-Mills fields (`letters'). The resulting physical picture is a covariant version of the BMN light-cone string, now with a finite number of twistorial string bit constituents of an essentially topological worldsheet. }

\begin{document}

\section{Introduction}

What does a stringy reformulation of Yang-Mills theory look like? We have long suspected, since 't~Hooft \cite{tHooft:1973alw}, that the large $N$ limit of gauge theories is the starting point for a dual string description. However, it was not until the advent of the AdS/CFT correspondence that we have had the conceptual apparatus to attempt an answer to the above question. The proposal \cite{Maldacena:1997re} for an ${\rm AdS}_5\times {\rm S}^5$ string dual to the interacting ${\cal N}=4$ super Yang-Mills theory in 4d was a very concrete one, especially when combined with the prescription \cite{Gubser:1998bc, Witten:1998qj} for relating string amplitudes to gauge invariant correlators. Among the many insights that this example has provided, a particularly liberating one was that the background (${\rm AdS}_5\times {\rm S}^5$) which the dual string sees is not the same as that in which the 4d gauge theory lives.   

One might have hoped that knowing what the dual to ${\cal N}=4$ super Yang-Mills (SYM) theory is, would immediately translate into uncovering the string duals to other Yang-Mills theories --- particularly those with no supersymmetry. After all, in a conventional QFT description, ${\cal N}=4$ SYM is nothing more than pure Yang-Mills (YM) theory with some special set of matter fields such that the beta function exactly vanishes due to a conveniently enhanced global symmetry. However, a transferance of insight, and generalisation of what we have learnt for SYM to ordinary gauge theories, has not yet happened. 

The reason can be traced back to our lack of fine grained understanding of the string dual to {\it weakly coupled} ${\cal N}=4$ super Yang-Mills theory. After all, it is in QFT perturbation theory that the distinction between SYM and ordinary YM seems relatively minor --- both are governed by their respective, not so different, free field UV fixed points. At strong coupling, asymptotic freedom in usual YM leads to a gapped theory which is quite different from the exactly conformal  ${\cal N}=4$ SYM. In the parlance of AdS/CFT,  the bulk dual in both cases have a similar, asymptotically ${\rm AdS}_5$ behaviour near the boundary, but the interior geometries (or IR) are quite different. While the background dual to ${\cal N}=4$ SYM is exactly ${\rm AdS}_5\times {\rm S}^5$, we do not know what that of free YM is. Therefore we have no first principles way of determining the deformation of the geometry away from the boundary for the pure YM case. We will only be able to understand how the operator corresponding to the YM interaction modifies the geometry if and when we have control over the string dual to the free YM theory. This is where our lack of control even over the string dual to the closely related free ${\cal N}=4$ SYM stymies us \cite{HaggiMani:2000ru, Sundborg:2000wp, Sezgin:2001zs, Mikhailov:2002bp, Sezgin:2002rt}. 
    
The hurdle is seemingly technical. The IIB string theory for a large radius ${\rm AdS}_5\times {\rm S}^5$ (and hence strongly coupled SYM) is described by type IIB supergravity with $\alpha'$ (higher derivative) corrections. However, the quantisation of a sigma model with an ${\rm AdS}_5\times {\rm S}^5$ target space (at arbitrary radius) comes up against numerous challenges. Firstly, the presence of Ramond-Ramond five-form flux rules out any easy RNS description, see e.g.\ \cite{Cho:2018nfn}. 
Secondly, the alternative Green-Schwarz description \cite{Metsaev:1998it} is fairly intractable except in a large radius expansion or certain pp-wave like limits, after fixing to a light-cone like gauge \cite{Metsaev:2000yf, Arutyunov:2009ga}. Thirdly, pure spinor descriptions \cite{Berkovits:2000fe}, which avoid some of the pitfalls of both the above approaches, have yet to be developed technically to the extent where they provide a calculational framework.\footnote{The pure spinor approach has, nevertheless, suggested various topological string descriptions \cite{Berkovits:2007zk, Berkovits:2007rj, Berkovits:2008qc} of the string dual to free Yang-Mills which share qualitatively similar features with our proposal in this paper.} This has frustrated efforts to describe the string theory beyond the large radius limit, let alone the highly curved regime which corresponds to weakly coupled ${\cal N}=4$ SYM. 
\smallskip

Recently, we have proposed a free field worldsheet sigma model that captures the tensionless (or zero radius) limit that describes free ${\cal N}=4$ super Yang-Mills \cite{Gaberdiel:2021iil}. It is somewhat orthogonal to the three approaches mentioned above, in that its starting point is not (directly) a sigma model on the ${\rm AdS}_5\times {\rm S}^5$ target space. The underlying geometry is, instead, a twistorial description of ${\rm AdS_5}\times {\rm S}^5$. It can also be equivalently viewed as a (super) ambitwistorial description of the 4d conformal boundary on which the Yang-Mills theory lives --- see e.g.\ \cite{Adamo:2016rtr}.  

The sigma model we propose in \cite{Gaberdiel:2021iil} is a direct progeny of a similar free field sigma model proposed for the tensionless limit of ${\rm AdS}_3\times {\rm S}^3\times {\mathbb T}^4$ a few years ago \cite{Eberhardt:2018ouy}. In that case the sigma model arose directly from the family of $\mathfrak{psu}(1,1|2)_k$ supergroup WZW models that describes ${\rm AdS}_3\times {\rm S}^3$ backgrounds (with $k$ units of NS-NS flux) in a hybrid formalism \cite{Berkovits:1999im}. The tensionless limit corresponds to the level $k=1$ and is where the supergroup WZW model admits a free field description \cite{Eberhardt:2018ouy}. This free field description is in terms of four free fermions (in two canonically conjugate pairs), together with a similar set of four symplectic bosons (two $\beta-\gamma$ pairs, each with spin one half). It can be thought of as a twistorial sigma model for the ${\rm AdS}_3\times {\rm S}^3$ background. The proposal has successfully reproduced the spectrum as well as correlation functions of the dual symmetric product orbifold CFT \cite{Eberhardt:2018ouy, Eberhardt:2019ywk, Eberhardt:2020akk, Dei:2020zui}. It therefore inspires confidence that the analogous model for ${\rm AdS_5}\times {\rm S}^5$ can capture the physics of free ${\cal N}=4$ SYM. 

In this paper, we will elaborate on certain aspects of our proposal in \cite{Gaberdiel:2021iil} and buttress the arguments in its favour. One of the most compelling pieces of evidence \cite{Gaberdiel:2021iil} was the agreement of the planar SYM operator spectrum with that of the string theory, using a certain physical gauge fixing prescription for the quantisation of the sigma model. Here we will derive this equivalence analytically, and in the process explain why the correspondence is structurally natural. In fact, we can think of our description as a covariant version of the BMN approach \cite{Berenstein:2002jq} in that the SYM spectrum is organised similarly but now with the $\mathfrak{psu}(2,2|4)$ global symmetry manifest. At the same time our worldsheet theory exhibits many qualitative features that have been part of the folklore for the tensionless limit of the dual string theory. We will also show how a similar physical gauge fixing in the case of the sigma model model for ${\rm AdS}_3\times {\rm S}^3$ gives rise to a spectrum which is a subsector of that of the dual symmetric product CFT. In some sense, it is the part of the spectrum which is independent of the compactification (e.g.\ to $\mathbb{T}^4$ or K3) of the original 10d string theory.   

\subsection{An extended outline}

Before we plunge into the technical details of the following sections it may be of some help to have a more granular takeaway picture of our results. 
Let us first recap the basic features of our proposed sigma model. It is essentially a doubled version of the one described above for ${\rm AdS}_3\times {\rm S}^3$. Namely, we now have eight real free fermions, i.e.\ four conjugate pairs $(\psi^a,\psi^\dagger_a)$, with $a\in\{1,2,3,4\}$. We also have eight free symplectic bosons, again organised in four conjugate pairs, namely  $(\lambda^\alpha, \mu^{\dagger}_{\alpha})$ and $(\mu^{\dot{\alpha}},\lambda^{\dagger}_{\dot{\alpha}})$, with $\alpha,\dot{\alpha} \in \{1,2\}$. Together they have a net vanishing central charge $c=0$. 
We often assemble them into a pair of ambitwistor variables, $Y_{I}= (\mu^\dagger_{\alpha},\lambda^\dagger_{\dot{\alpha}},\psi^{\dagger}_a)$ and 
$Z^{I}=(\lambda^\alpha,\mu^{\dot{\alpha}},\psi^a)$. They obey the ambitwistor constraint 
\be\label{Cdef}
{\cal C} =\tfrac{1}{2} Y_IZ^I =0 \ ,
\ee
which will be implemented by demanding that the states of our theory are annihilated by the positive modes of ${\cal C}$. 
The vanishing of this `central' term ensures that we obtain a free field realisation of the $\mathfrak{psu}(2,2|4)_1$ current algebra (rather than $\mathfrak{u}(2,2|4)$). Note that all of these fields are left-movers, and there is a corresponding set of right movers as well. 

In ${\rm AdS}_3$ all the physics came from considering spectrally flowed representations \cite{Maldacena:2000hw}, and it will be the same in the present instance too. 
Spectral flow is an automorphism of the current algebra which takes highest weight affine representations to other representations which are generically not highest weight. In our case, it acts by shifting the mode numbers of the free fields up or down by units of $\frac{w}{2}$ --- see eq.~\eqref{spectralflow}, where $w >0$ is an integer that labels the amount of spectral flow. 

The action of spectral flow can be viewed as one in which the field modes have a modified behaviour when acting on the spectrally flowed vacuum state, which we denote by $|0\rangle_w$ in the sector with $w$ units of spectral flow. The modes which have a non-zero action on $|0\rangle_w$ fall into two broad categories: 
\begin{list}{(\roman{enumi})}{\usecounter{enumi}}
\item The {\it wedge modes}, i.e.\
\be\label{wedge0}
(\mu^\dagger_\alpha)_r \ , \quad (\mu^{\dot{\alpha}})_r \ , \quad   
(\psi^\dagger_{a})_r \  \ (a=1,2) \ ,  \quad \psi^{b}_r \ \ (b=3,4)\ ,  \qquad (-\tfrac{w-1}{2} \leq r \leq \tfrac{w-1}{2})
\ee
\item The {\it out-of-the-wedge modes}, i.e.\
\be\label{oowedge}
Z^I_r \ \hbox{and} \ (Y_J)_r  \qquad \hbox{with $r\leq - \frac{w+1}{2}$} \ .
\ee
\end{list}

Note that the `wedge' modes involve only half the fields of the $Y_I$ and $Z^I$. 
We will view the wedge modes as generalised zero modes for each spectrally flowed sector. In the context of the twistor string theory of \cite{Berkovits:2004hg}, these are the analogues of the zero modes in the topological sector with $w$ units of $\mathfrak{u}(1)$ flux, corresponding to the current ${\cal C}$ in eq.~\eqref{Cdef}. The `out-of-the-wedge' modes are then the analogues of the non-zero modes. 

In fact, for $w=1$ the wedge modes are indeed just the zero modes, and the resulting Fock space generated by these modes gives the usual oscillator construction \cite{Gunaydin:1984fk, Beisert:2004ry} which generates representations of 
$\mathfrak{psu}(2,2|4)$ starting with the so-called singleton. The latter will be an important building block in our construction. 
  
One of our central claims is that there is a quantisation of this sigma model in which we can pick an especially natural gauge for the underlying worldsheet constraints. In this `physical' gauge all the 
out-of-the-wedge modes can be set to zero, leaving only finitely many oscillators, for each $w$, namely the wedge modes in eq.~\eqref{wedge0}. We will dub this the `topological' gauge since only the generalised zero modes survive. We will view these wedge modes as exciting a discrete set of $w$ string bits which are localised along the worldsheet, where the localised `position space' generators $\hat{\Phi}_j$ are defined via 
\be\label{Four}
\hat{\Phi}_j = \frac{1}{\sqrt{w}}\, \sum_{r=- (w-1)/2}^{(w-1)/2} \Phi_r \, e^{-2\pi i\frac{rj}{w}}  \ ,
\ee
see Section~\ref{sec:4.1} for more details.  This is very much like the BMN picture \cite{Berenstein:2002jq}, the difference being that here we have twistorial degrees of freedom $\hat{\Phi}_j$ at each site. The usual light-cone oscillators of BMN are composites built from these twistor bits.\footnote{A string bit picture in light cone gauge, dual to the free SYM theory and generalising the BMN analysis, was explored in \cite{Karch:2002vn, Dhar:2003fi, polch2002}. In the language of condensed matter physics, what we have here is a fractionalisation of those degrees of freedom.} Since the out-of-the-wedge modes are not physical, the bulk of the worldsheet is non-dynamical or topological. In addition, we would like to claim that for $w>0$ there are no physical right movers either. Again, this has an analogue in the usual twistor string theory where only holomorphic sections are present in a line bundle with a definite sign of the $\mathfrak{u}(1)$ flux. 

On the Fock space generated by these wedge oscillators, we still need to impose some residual gauge constraints. We claim that the physical states $|\Psi_{\rm phy}\rangle \in {\cal H}_{\rm phys}$ are characterised by the two conditions:
\begin{list}{(\alph{enumi})}{\usecounter{enumi}}
\item The residual Virasoro constraint 
\be\label{lzero}
(L_0+pw)\, |\Psi_{\rm phy}\rangle =0  \qquad (p\in \mathbb{Z} )
\ee
\item The `central' term constraint
\be\label{cn}
{\cal C}_n \, |\Psi_{\rm phy}\rangle =0 \qquad (n=0,1,\ldots , w-1) \ . 
\ee  
\end{list}
Here $L_0$ counts the oscillator mode numbers in the usual way, while the different modes have charges $\pm \frac{1}{2}$ with respect to ${\cal C}_n$, i.e.\ 
$[{\cal C}_n, \Phi_r]=\pm \frac{1}{2}\Phi_{n+r}$; the modes of $\mu^\dagger_{\alpha}$ and $\psi^{\dagger}_{1,2}$ 
carry $(+)$ charge with respect to ${\cal C}_n$, while those of $\mu^{\dot{\alpha}}$ and $\psi^{3,4}$  carry $(-)$ charge. Also $\Phi_{m}\equiv 0$  if $m > \frac{w-1}{2}$.  

We can think of the first constraint (a) as imposing momentum conservation on the $w$ discrete sites which comprise the string bit picture of the physical worldsheet. Because of the discreteness, momentum is only conserved modulo $w$, as in any lattice theory. From this perspective we can view eq.~\eqref{lzero} as imposing the cyclicity in position space along the worldsheet, see eq.~(\ref{L0}) below. This constraint can also be viewed as analogous to the residual Virasoro constraint in light-cone gauge where $L_0^{\perp} \propto p^+$. Here the light-cone momentum characterises the ground state in a given sector, and we identify $w \sim p^+$. This is a natural identification since, as we will see, $|0\rangle_w$ corresponds to the half BPS state with quantum numbers $[0,w,0]$ under $\mathfrak{su}(4)$. In other words, this state should be identified (for large $w$) with the BMN vacuum state in which $p^+ \propto w$. There is also a sense in which the integer $p$ plays the role of $2p^-$, see eq.~\eqref{oscnum} below. 

The second set of constraints (b) are what remains from the requirement of eq.~\eqref{Cdef}, which we need to impose in the original sigma model to go from $\mathfrak{u}(2,2|4)$ to $\mathfrak{psu}(2,2|4)$. There are $w$ such constraints, and they turn out to demand that the wedge oscillators give rise to a single copy of the singleton representation of $\mathfrak{psu}(2,2|4)$ at each of the $w$ sites, see eq.~(\ref{Ccond}) below. Thus the two constraints together imply that the space of physical states is the $w$-fold tensor product of the singleton representation, subject to a cyclicity condition. 

This is the underlying reason why the physical string Hilbert space, for any fixed $w$ as described above, is in one-to-one correspondence with operators of planar SYM built from $w$ copies of the SYM fields (and their derivatives). Recall that the singleton representation consists of a single copy (`letter') of the SYM fields, $S= \{ \partial^s\phi^i, \partial^s\Psi^a_{\alpha}, \partial^s\Psi_a^{\dot{\alpha}}, \partial^s{\cal F}_{\alpha, \beta}, \partial^s {\cal F}^{\dot{\alpha}, \dot{\beta}} \}$, modulo the equations of motion of the free theory. The planar spectrum of ${\cal N}=4$ SYM then consists of single trace operators or `words' built from $w$ letters, i.e\ they are of the form ${\rm Tr}\bigl(S_1, S_2\cdots S_w\bigr)$ \cite{Sundborg:1999ue, Polyakov:2001af, Bianchi:2003wx,Aharony:2003sx}. Thus it also arises from the tensor product of $w$ singletons subject to cyclicity under the trace. The worldsheet can then be viewed as $w$ beads on a necklace (as in the Polya counting!), with the beads living at discrete sites and having dynamical degrees of freedom. The rest of the worldsheet (the chain) which connects the beads  is rigid or topological with those modes being pure gauge (on the worldsheet). 

This structural mapping between the physical spectrum generated by the wedge mode oscillators of our sigma model, and the planar spectrum of SYM is one of the strong reasons to believe in this worldsheet description. At a microscopic level, the worldsheet picture gives us a novel way of organising the SYM spectrum, now stratified in terms of the eigenvalues of $L_0$ in eq.~\eqref{lzero}. As mentioned, this is the covariant analogue of the BMN light-cone Hamiltonian (at zero coupling). We believe this may be a fruitful way of thinking about the SYM spectrum when one goes away from zero coupling, thereby lifting the large degeneracy of states of the free theory. 

We will therefore describe in some detail the low lying SYM spectrum as organised by the worldsheet theory, see Section~\ref{sec:spectrum}. In doing so we observe that all physical states can be generated from the ground state by a set of DDF-like operators, see eq.~(\ref{Sdef}). These have an interesting algebraic structure and obey several non-trivial relations which it would be interesting to explore further. A simple set of operators which manifestly match for all $w$ are the half-BPS multiplets built from the highest weight state carrying the representation $(0,0; [0,w, 0])$ under the $\mathfrak{su}(2)\oplus\mathfrak{su}(2)\oplus \mathfrak{su}(4)$ subalgebra of $\mathfrak{psu}(2,2|4)$. These arise on the worldsheet from the spectrally flowed vacuum $|0\rangle_w$ and its $\mathfrak{psu}(2,2|4)$ descendants, and all of them have $L_0=0$. In \cite{Gaberdiel:2021iil} we had already studied the $w=2$ case in some detail,  and showed how the different higher spin fields (twist two operators) correspond to the worldsheet states with $L_0=-2p$, see also eq.~(\ref{hspin}) below. Here we go on to consider the $w=3$ case  and enumerate the representations that match with those of the super Yang-Mills theory. 
\smallskip

Our matching with the SYM spectrum rests on the proposal for what the physical degrees of freedom  are in our `topological gauge'. Such a gauge fixing procedure should also apply to the case of ${\rm AdS}_3\times {\rm S}^3$,  although that is not how we had matched the spectrum in that case with the dual CFT. One consistency check is therefore to revisit the spectrum of the tensionless string theory on  ${\rm AdS}_3\times {\rm S}^3$ by looking at the analogue of the wedge modes of eq.~\eqref{wedge0} in that case, and imposing the same physical state conditions on their Fock space, see conditions (a) and (b) above. 

Again we can count  the physical states in terms of cyclically invariant combinations of tensor powers of the `singleton' representation of $\mathfrak{psu}(1,1|2)$. (We have also confirmed this explicitly by brute force, at least for low levels.) We will show that these states can be identified with a  certain natural subset of states in the symmetric orbifold theory of $\mathbb{M}=\mathbb{T}^4$ or $\mathbb{M} =$ K3 --- since our analysis only sees the ${\rm AdS}_3\times {\rm S}^3$ part of the background, we should not expect to reproduce all physical states, nor should the answer depend on whether we consider $\mathbb{T}^4$ or K3. The physical states we find  include the even spin higher spin fields (that are present in both theories), as well as a half-BPS multiplet from each $w$-cycle twisted sector. In fact, we find the BPS multiplet that corresponds to the  zeroth cohomology of $\mathbb{M}$ --- recall that in any maximally supersymmetric compactification ${\rm AdS}_3\times {\rm S}^3\times \mathbb{M}$, there are half BPS multiplets in correspondence with the cohomology elements of $\mathbb{M}$. This obviously makes a lot of sense since the zeroth cohomology of $\mathbb{M}$ is independent of the form of  $\mathbb{M}$,  as befits our worldsheet analysis that does not directly involve $\mathbb{M}$. We take this as additionally bolstering the case that our twistorial sigma models are essentially topological and have only wedge modes as their physical modes.  
\smallskip

Up to now we have focussed on discussing the consequences of assuming a physical gauge fixing that leaves us with only the wedge modes, together with the residual constraints. We also give a short discussion on the pathway to quantising our original sigma model which would enable us to derive these as the physical modes, see Section~\ref{Sec:3}. This is again guided by our experience with ${\rm AdS}_3\times {\rm S}^3\times {\mathbb T}^4$. The fact that the theory on the ${\rm AdS}_3\times {\rm S}^3$ factor is essentially topological  can be seen in the microscopic analysis from the fact that there is an underlying worldsheet (twisted) $N=2$ gauged superconformal invariance. This was exploited by Berkovits, Vafa and Witten to quantise the $\mathfrak{psu}(1,1|2)_k$ sigma model (for general $k$) \cite{Berkovits:1999im}. For the tensionless $k=1$ theory with the free field realisation, these constraints of the theory are as many as the free fermion and boson degrees of freedom, leaving only topological or generalised zero modes which are the wedge modes. 

In the present instance, with eight free bosons and fermions, we can construct more superconformal generators on the worldsheet and it is natural to expect that there is an enhanced gauged $N=4$ worldsheet superconformal invariance. If this were to be the case then  the corresponding constraints would remove all the degrees of freedom, leaving at most the generalised zero modes or wedge modes as topological degrees of freedom. In this context, we note that the matter as well as the (twisted) $N=4$ ghosts separately form $c=0$ systems as needed for a critical string theory. 
\medskip

The paper is organised as follows. In Section~\ref{Sec:1} we describe the twistorial worldsheet fields as well as the action of spectral flow which gives rise to a natural decomposition into wedge and 
out-of-the-wedge modes. In Section~\ref{Sec:3} we sketch how this sigma model might be quantised in analogy to the ${\rm AdS}_3\times {\rm S}^3$ case, and how this ought to lead to a gauge fixed description involving the wedge modes together with some residual constraints.  Assuming this can be done, we study the physical spectrum that is generated by these wedge oscillators, and demonstrate its matching with that of free super Yang-Mills theory at the planar level. We then go on to make a more refined analysis of the spectrum as organised by the worldsheet constraints; this is done in Section~\ref{sec:spectrum}, see also Appendix~\ref{app:B}. In Section~\ref{sec:AdS3} and Appendix~\ref{app:C} we carry out a similar analysis  for the case of ${\rm AdS}_3\times {\rm S}^3$. Finally, we end with some concluding remarks in Section~\ref{sec:concl}. In addition to the appendices mentioned already above, we have summarised our notation and the free field realisation of the $\mathfrak{psu}(2,2|4)$ algebra in Appendix~\ref{app:psu}.

\section{The free field worldsheet theory}\label{Sec:1}
In this section we will describe the worldsheet theory of free fermions and bosons, together with its spectrally flowed sectors, which lies at the heart of our proposal. After describing the construction of the generators of the global $\mathfrak{psu}(2,2|4)$ symmetry from the free fields, we spell out the action of the spectral flow starting from the NS vacuum. 

Our proposal for the worldsheet theory dual to free super Yang-Mills in d=4 is inspired by the free field realisation of $\mathfrak{psu}(1,1|2)_1$ that played a key role in the ${\rm AdS}_3 / {\rm CFT}_2$ duality of \cite{Eberhardt:2018ouy}. The super Lie algebra $\mathfrak{psu}(1,1|2)_1$ has a free field realisation in terms of two pairs of symplectic bosons (i.e.\ a first order system of bosons of spin half, see eq.~\eqref{ffcomm} below) and two pairs of complex fermions, modulo two $\mathfrak{u}(1)$ fields, see \cite{Dei:2020zui} and Appendix~\ref{app:AdS3} for more details. In the present setup we basically just double these degrees of freedom; this then leads to a free field realisation of $\mathfrak{psu}(2,2|4)_1$, and hence guarantees that our worldsheet theory possesses the superconformal $\mathfrak{psu}(2,2|4)$ symmetry of the dual SYM theory. More specifically, we consider two pairs of (pairs of) symplectic boson fields $(\lambda^\alpha,\mu^{\dagger}_{\alpha})$ and $(\mu^{\dot{\alpha}},\lambda^{\dagger}_{\dot{\alpha}})$ with $\alpha,\dot{\alpha} \in \{1,2\}$, as well as four complex fermions $(\psi^a,\psi^\dagger_a)$ with $a\in\{1,2,3,4\}$, and commutation relations 
\be\label{ffcomm}
[\lambda^\alpha_r,(\mu^\dagger_\beta)_s] = \delta^{\alpha}_{\beta} \, \delta_{r,-s} \ , \qquad 
[\mu^{\dot{\alpha}}_r,(\lambda^\dagger_{\dot{\beta}})_s] = \delta^{\dot{\alpha}}_{\dot{\beta}} \, \delta_{r,-s} \ , \qquad 
\{\psi^a_r,(\psi^\dagger_b)_s \} = \delta^{a}_b \, \delta_{r,-s} \ . 
\ee
We combine the oscillators as $Y_ J= (\mu^\dagger_{\alpha},\lambda^\dagger_{\dot{\alpha}},\psi^\dagger_a)$ and $Z^I = (\lambda^{\alpha},\mu^{\dot{\alpha}},\psi^a)$, and then consider the bilinears 
\be
J^I{}_J = Y_J \, Z^I \ .
\ee
It follows essentially from the analysis in \cite[Appendix~D]{Beisert:2004ry} that these fields generate the current algebra corresponding to $\mathfrak{u}(2,2|4)$; this is just the current algebra version of the more familiar oscillator construction of $\mathfrak{u}(2,2|4)$ which is the building block of the on-shell physical representations in Yang-Mills theory \cite{Gunaydin:1984fk}, and which will play an important role in what follows. Explicit expressions for the different generators and their commutation relations can be found in Appendix~\ref{app:psu}; we also spell out our notation and conventions there.

One of the generators of $\mathfrak{u}(2,2|4)$ that will play a critical role is ${\cal C}= Y_I \, Z^I$.
Its modes ${\cal C}_n$ are central (except for one non-trivial commutator with another $\mathfrak{u}(1)$ generator ${\cal B}_n$, see eq.~(\ref{BC})). It needs to be `modded out' in order to go to  $\mathfrak{psu}(2,2|4)_1$. As in \cite{Dei:2020zui} this will be implemented by working with the subspace on which ${\cal C}_n$ with $n\geq 0$ annihilates states; the ${\cal C}_{-n}$ descendants are then null, and are thus naturally quotiented out.

\subsection{The spectrum}

Having identified the free fields of the worldsheet description, the next step is to specify which representations appear in the worldsheet theory. The simplest representation of $\mathfrak{psu}(2,2|4)_1$ arises in the NS sector where all the modes $Z^I_r$ and $(Y_J)_r$ are half-integer moded. On the ground state of the NS sector we have the eigenvalues for the generators $U_0 \, |0\rangle  = \dot{U}_0 \, |0\rangle=0$, as well as $V_0 \, |0\rangle=0$ (see eq.~\eqref{UVdef}). The same is also true for the Cartan generators of $\mathfrak{su}(2) \oplus \mathfrak{su}(2) \oplus \mathfrak{su}(4)$, as well as  
\be\label{NSeigenvals}
{\cal B}_0 \, |0\rangle =  {\cal C}_0 \, |0\rangle =  {\cal D}_0\, |0\rangle = 0  \ .
\ee
While for ${\cal B}_0$ this is a matter of convention (albeit the natural one), this is imposed for ${\cal C}_0$ and ${\cal D}_0$ by the (anti)-commutators of eqs.~(\ref{KP}) -- (\ref{dotSdotQ}). 
This therefore defines the usual vacuum representation of $\mathfrak{psu}(2,2|4)_1$, and we shall denote it by ${\cal H}_0$ in the following. It  will turn out to contain only one physical state, namely the vacuum state itself, which will correspond to the identity operator of the super Yang-Mills theory.

\subsection{Spectral flow}

The NS-sector representation (as well as the R sector representation, which we will discuss in the next subsection) are the only highest weight representations of this free field theory, but, as for the case of ${\rm AdS}_3$, we expect the spectrum to contain in addition representations that are obtained by {\bf spectral flow} from these highest weight representations \cite{Henningson:1991jc,Maldacena:2000hw}.\footnote{One difference from the case of ${\rm AdS}_3$ that will be useful to keep in mind is that here we will find it more natural to measure the spectral flow with respect to the NS-representation. In the case of ${\rm AdS}_3$ spectral flow $w$ is usually measured with respect to the Ramond representation. The two conventions for the parameter $w$  are therefore shifted by $1$ relative to one another --- see eq.~\eqref{shift} below.} To describe this in detail we define the $w$-spectrally flowed modes (with tildes) via 
\begin{equation} \label{spectralflow}
\begin{array}{rclrcl}
(\tilde{\lambda}^\alpha)_r & = & (\lambda^\alpha)_{r-w/2}  \ ,  \qquad  \qquad 
& (\tilde{\lambda}^\dagger_{\dot{\alpha}})_r & = & (\lambda^\dagger_{\dot{\alpha}})_{r-w/2} \ ,   \\[4pt] 
(\tilde{\mu}^{\dot{\alpha}})_r  & =  & (\mu^{\dot{\alpha}})_{r+w/2}  \ , \qquad \qquad 
& (\tilde{\mu}^\dagger_\alpha)_r  & =  & (\mu^\dagger_\alpha)_{r+w/2}  \ ,   \\[4pt]
(\tilde{\psi}^{a}_r) & = & \psi^a_{r-w/2} \ , \qquad  \qquad 
& (\tilde{\psi}^\dagger_{a})_r & = & (\psi^\dagger_{a})_{r+w/2}  \qquad \qquad  (a=1,2) \ ,   \\[4pt]
(\tilde{\psi}^{b}_r) & =  & \psi^b_{r+w/2} \ , \qquad  \qquad
& (\tilde{\psi}^\dagger_{b})_r & =  & (\psi^\dagger_{b})_{r-w/2}     \qquad \qquad  (b=3,4) \ . 
\end{array}
\end{equation}
The $w$-spectrally flowed representation is then defined by letting the tilde modes act on the original NS-sector highest weight representation --- so in particular, the positive tilde modes annihilate  the highest weight states --- but interpreting the representation in terms of the original modes (without tildes). For example, the $w$-spectrally flowed image of the NS ground state, which we shall denote by $|0\rangle_w$ in the following, is then characterised by the property that the modes 
\begin{align}
& \mu^{\dot{\alpha}}_r\, |0\rangle_w  =   (\mu^\dagger_\alpha)_r \, |0\rangle_w  =   (\psi^\dagger_{1,2})_{r}   \,  |0\rangle_w  = \ \psi^{3,4}_{r} \, |0\rangle_w  =0 \ , \qquad  &( r \geq \tfrac{w+1}{2} )  \label{prop1} \\
 & \lambda^\alpha_r\,  |0\rangle_w  =  (\lambda^\dagger_{\dot{\alpha}})_r\,  |0\rangle_w  = (\psi^{1,2})_r \,  |0\rangle_w  =  (\psi^{\dagger}_{3,4})_r \,  |0\rangle_w  = 0 \ ,  \qquad  & ( r \geq - \tfrac{w-1}{2} ) \label{prop2}
 \end{align}
annihilate $|0\rangle_w$. (In this sector, the mode numbers are of the form 
$r \in \frac{w-1}{2} + \mathbb{Z}$.) Conversely, the modes 
\begin{align}\label{nontriv1}
& \mu^{\dot{\alpha}}_r\ , \  (\mu^\dagger_\alpha)_r\ ,  \  (\psi^\dagger_{1,2})_{r}  \ ,  \ \psi^{3,4}_{r} \ , \qquad  \qquad & ( r \leq \tfrac{w-1}{2} ) \\
& \lambda^\alpha_r\ , \ (\lambda^\dagger_{\dot{\alpha}})_r\ , \ (\psi^{1,2})_r \ , \  (\psi^{\dagger}_{3,4})_r \ , \qquad  \qquad &  ( r \leq - \tfrac{w+1}{2} ) \nonumber
 \end{align}
 then freely generate the full Fock space from $|0\rangle_w$. It is convenient to organise the above modes into the wedge modes and out-of-the-wedge modes as in eqs.~\eqref{wedge0} and \eqref{oowedge}, respectively.
 
 Our definition of spectral flow in 
 eq.~\eqref{spectralflow} is motivated by the requirement that the spacetime translation operator ${\cal P}^{\dot{\alpha}}{}_{\beta} = \mu^{\dot{\alpha}} \mu^\dagger_{\beta}$, which is the 4d analogue of the generator $J^+_0 \in \mathfrak{sl}(2,\mathds{R})$, should transform as 
\be
(\tilde{{\cal P}}^{\dot{\alpha}}{}_{\beta})_{r}= ({\cal P}^{\dot{\alpha}}{}_{\beta})_{r+w} \ ,
\ee
so that its zero mode $({\cal P}^{\dot{\alpha}}{}_{\beta})_0 = (\tilde{{\cal P}}^{\dot{\alpha}}{}_{\beta})_{-w}$ acts non-trivially on $|0\rangle_w$, cf.\ with the situation for ${\rm AdS}_3$ in \cite{Eberhardt:2018ouy,Dei:2020zui};\footnote{Relative to \cite{Eberhardt:2018ouy,Dei:2020zui}, where spectral flow was described in terms of an `active' automorphism $\sigma$, we are taking here the `passive' perspective. In the notation of \cite{Eberhardt:2018ouy,Dei:2020zui}, $|0\rangle_w = [|0\rangle]^{(w)}$, with action $\lambda^\alpha_r [|0\rangle]^{(w)} = [ \sigma^w(\lambda^\alpha_r)|0\rangle]^{(w)}$, so the above convention of spectral flow corresponds to $\sigma^w(\lambda^\alpha_r)=\lambda^\alpha_{r+\frac{w}{2}}$, and similarly for the other modes.\label{foot1}}
 this fixes the spectral flow action on the symplectic bosons. Then the $\mathfrak{su}(2) \oplus \mathfrak{su}(2)$ generators ${\cal L}^\alpha{}_\beta$ and $\dot{\cal L}^{\dot{\alpha}}{}_{\dot{\beta}}$ are spectral flow invariant, while the $\mathfrak{u}(1)$ current modes transform as 
\be
\tilde{U}_n = U_n - w \, \delta_{n,0} \ , \qquad 
\tilde{\dot{U}}_n = \dot{U}_n + w \, \delta_{n,0} \ , \qquad 
\tilde{V}_n = V_n \ , 
\ee
where we have required $V_n$ to transform trivially so that 
\be\label{tildeC}
\tilde{{\cal B}}_n = {\cal B}_n \ , \qquad \tilde{{\cal C}}_n = {\cal C}_n \ .
\ee
This is demanded by consistency so that representations of $\mathfrak{psu}(2,2|4)$ (for which we have ${\cal C}_0=0$) are mapped onto one another under spectral flow. This then fixes the spectral flow action on the fermions, up to an automorphism by $\mathfrak{su}(4)$. With these conventions we then find that the spacetime scaling (dilatation) operator transforms as 
\be\label{DRspecflow}
\widetilde{\cal D}_n = {\cal D}_n - w\, \delta_{n,0} \ ,
\ee
which is the natural analogue of how $J^3_n \in \mathfrak{sl}(2,\mathds{R})$ transforms for the case of ${\rm AdS}_3$.  We have chosen our conventions in eq.~(\ref{spectralflow}) such that the spectral flow automorphism acts on the diagonal $\mathfrak{u}(4)$ generators ${\cal R}^{a}{}_a$ as 
\be\label{Rsf}
(\widetilde{{\cal R}}^{a}{}_a)_n  = \left\{ 
\begin{array}{cl}
({\cal R}^{a}{}_a)_{n} + \tfrac{w}{2} \delta_{n,0} \qquad & \hbox{if $a=1,2$} \\[4pt] 
({\cal R}^{a}{}_a)_{n} - \tfrac{w}{2} \delta_{n,0} \qquad & \hbox{if $a=3,4$,} 
\end{array}
\right.
\ee
while the off-diagonal generators ${\cal R}^{a}{}_b$ with $a\neq b$ transform as  
\be\label{Rsf1}
(\widetilde{{\cal R}}^{a}{}_b)_n  = \left\{
\begin{array}{cl}
({\cal R}^{a}{}_b)_{n+w}  \qquad & \hbox{if $b\in \{1,2\}$ and $a\in\{3,4\}$,} \\
({\cal R}^{a}{}_b)_{n-w}  \qquad & \hbox{if $a\in \{1,2\}$ and $b\in\{3,4\}$,} \\
({\cal R}^{a}{}_b)_{n}  \qquad & \hbox{if $a,b\in \{1,2\}$ or $a,b\in\{3,4\}$.} 
\end{array}
\right.
\ee
Spectral flow then acts on the Cartan generators of $\mathfrak{su}(4)$, defined in eq.~\eqref{cartan}, as
\be\label{Hiflow}
(\tilde{H}_1)_n = (H_1)_n \ , \qquad (\tilde{H}_2)_n  = (H_2)_n - w\, \delta_{n,0}  \ , \qquad (\tilde{H}_3)_n = (H_3)_n \ . 
\ee
Finally, the worldsheet Virasoro algebra transforms as
\be\label{tildeL}
\tilde{L}_n = L_n - w({\cal D}_n - {\cal R}_n) \ ,
\ee
where ${\cal R}_n$ is the combination 
\be
{\cal R}_n= \frac{1}{2}\Bigl[ - ({\cal R}^1{}_1)_n - ({\cal R}^2{}_2)_n  +  ({\cal R}^3{}_3)_n +  ({\cal R}^4{}_4)_n \Bigr] \ , 
\ee
for which we have 
\be
\widetilde{\cal R}_n = {\cal R}_n - w\, \delta_{n,0} \ .
\ee

The full spectrum of the worldsheet theory consists then of the NS sector representation (both for left- and right-movers), together with all its spectrally flowed images, where we only consider $w\geq 0$, and let the spectral flow act by the same amount on left- and right-movers. This is the natural analogue of the worldsheet theory for $\mathfrak{psu}(1,1|2)_1$ from \cite{Eberhardt:2018ouy}, see also \cite{Maldacena:2000hw}. 

\subsection{Ramond representation}\label{sec:Ramond}

One may wonder how the Ramond representation fits into this general framework. In this subsection we shall define it directly; we shall see in the next subsection that it is actually equivalent to the $w=1$ spectrally flowed NS representation.

In the Ramond sector, all modes $Z^I_n$ and $(Y_J)_n$ are integer moded, and we therefore have to be careful about the correct normal ordering prescription, in particular in the definition of ${\cal C}_0$ and ${\cal D}_0$. We label the states by the symplectic boson occupation numbers $|m_1,m_2;\dot{m}_1,\dot{m}_2\rangle$, where the action of the zero modes are given in (\ref{laaction}) and  (\ref{muaction}). The eigenvalues of the Cartan generators of $\mathfrak{u}(2,2|4)$ follow from this, and the explicit results are given in Appendix~\ref{app:R}. 

In the $\mathfrak{psu}(2,2|4)$ symmetry of the dual CFT, the $\mathfrak{su}(2) \oplus \mathfrak{su}(2)$ subalgebra  is compact, and hence the spins (and the $J^3_0$ and $\hat{J}^3_0$ eigenvalues) must be half-integers. Because of (\ref{J3eig}) and (\ref{J3heig}), as well as  (\ref{Ceig}) and (\ref{Cheig}), this requires that 
\be
m_1,m_2 \in \tfrac{1}{2} \mathbb{Z} \ , \qquad \dot{m}_1,\dot{m}_2 \in \tfrac{1}{2} \mathbb{Z} \ , \qquad m_1,\dot{m}_2 \in \tfrac{1}{2}\mathbb{N}_0 \ . 
\ee 
The quadratic Casimir of $\mathfrak{psu}(2,2|4)$ vanishes on these states, see eq.~(\ref{Cas}), and thus the representation is in fact indecomposable; the 
subspace generated by the vectors with 
\be\label{indecomp}
m_1,\dot{m}_1, m_2, \dot{m}_2 \in \tfrac{1}{2}\mathbb{N}_0
\ee
forms an (irreducible) subrepresentation (see eqs.~(\ref{laaction}) and (\ref{muaction})), and we shall in the following concentrate on this subspace. 

This highest weight space is annihilated by the modes $Z^I_n$ and $(Y_J)_n$ with $n>0$, and the full affine representation is generated from it by the action of the non-positive modes. The zero mode action of the symplectic bosons is already encapsulated by the above occupation numbers, but we also have the action of the fermionic zero modes. They generate a Clifford algebra representation, and with respect to the $\mathfrak{su}(4)$ generators (that are bilinears in the fermions) the states for fixed values of $m_i$ and $\dot{m}_i$ transform as 
\be
[0,1,0] \oplus  [1,0,0] \oplus  [0,0,1] \oplus 2 \cdot [0,0,0] \ , 
\ee
where  $[n_1,n_2,n_3]$ are the Dynkin labels  of $\mathfrak{su}(4)$, and we use the convention that $[1,0,0]\cong {\bf 4}$, $[0,0,1] \cong \overline{\bf 4}$, while $[0,1,0] \cong {\bf 6}$ is the adjoint representation. The values of $m_i$ and $\dot{m}_i$ only affect the eigenvalues with respect to $\mathfrak{su}(2) \oplus \mathfrak{su}(2)$, as well as the value of the dilatation operator ${\cal D}_0$, and the full highest weight representation therefore has the structure 
\begin{align}
{\cal R}_0 =&  \bigoplus_{s=0}^{\infty}  \Bigl[ \bigl( \tfrac{s}{2},\tfrac{s}{2};[0,1,0] \bigr)_{1+s} 
\oplus \bigl( \tfrac{s}{2}+1,\tfrac{s}{2} ; [0,0,0]\bigr)_{2+s}  \oplus
\bigl( \tfrac{s}{2},\tfrac{s}{2}+1 ; [0,0,0]\bigr)_{2+s} \nonumber \\
& \qquad \oplus 
\bigl( \tfrac{s+1}{2},\tfrac{s}{2}; [0,0,1] \bigr)_{\frac{3}{2}+s} \oplus
\bigl( \tfrac{s}{2},\tfrac{s+1}{2} ; [1,0,0] \bigr)_{\frac{3}{2}+s} \Bigr] \ ,  \label{singleton}
\end{align}
where $(j,\hat{\jmath}; [n_1,n_2,n_3])_s$ labels the states for which $j$ and $\hat{\jmath}$ are the spins with respect to the two $\mathfrak{su}(2)$ algebras, while $s$ is the eigenvalue of ${\cal D}_0$. (Note that we have used here eq.~\eqref{indecomp} which guarantees that the $\mathfrak{su}(2)$ spins are all half-integer.) This is the so-called {\bf singleton representation} of $\mathfrak{psu}(2,2|4)$, and it will play an important role in the following. As mentioned in the introduction, this representation is in one-to-one correspondence with the basic letters of the SYM theory, namely the set of SYM fields $\{ \partial^s\phi^i, \partial^s\Psi^a_{\alpha}, \partial^s\Psi_a^{\dot{\alpha}}, \partial^s{\cal F}_{\alpha, \beta}, \partial^s {\cal F}^{\dot{\alpha}, \dot{\beta}} \}$.  

We shall use round brackets to denote representations of the  bosonic subalgebra $\mathfrak{su}(2)\oplus \mathfrak{su}(2) \oplus \mathfrak{su}(4)$, and straight brackets for the representations of $\mathfrak{psu}(2,2|4)$. Since the singleton representation is irreducible with respect to $\mathfrak{psu}(2,2|4)$ with highest weight 
$\bigl( 0,0;[0,1,0] \bigr)_{1}$ we shall therefore denote it as 
\be\label{singleton1}
{\cal R}_0 \equiv \bigl[ 0,0;[0,1,0] \bigr]_{1} \ . 
\ee
The full affine R-sector representation, generated by the modes with $n<0$, (whose highest weight space is the singleton representation ${\cal R}_0$) will be denoted by ${\cal R}$ in the following.

\subsection{The spectral flow of the NS representation}

As mentioned above, the R sector representation ${\cal R}$ is in fact the image of the NS sector highest weight representation under one unit of spectral flow. More explicitly 
\be\label{NStoR}
|0\rangle_{w=1} =    (\psi^\dagger_3)_0  (\psi^\dagger_4)_0 |0,0;0,0\rangle \ .
\ee
In order to prove this we note that all positive modes, as well as the zero modes $(\lambda^\alpha)_0$, $(\lambda^\dagger_{\dot{\alpha}})_0$, $\psi^a_0$ with $a=1,2$, and $(\psi^\dagger_b)_0$ with $b=3,4$ annihilate both sides. Furthermore, one checks that also the eigenvalues of all the Cartan generators agree. 

Thus the $w$'th spectrally flowed image of the NS sector representation ${\cal H}_0$ can be viewed as the $(w-1)$'th spectrally flowed image of ${\cal R}$. For the following we shall find it more convenient to use the former description, and to characterise the  resulting representation directly: it is generated from the `vacuum' state $|0\rangle_w $ by the action of the modes 
\be\label{wedge}
(\mu^\dagger_\alpha)_r \ , \quad (\mu^{\dot{\alpha}})_r \ , \quad 
(\psi^\dagger_{a})_r \ \ (a=1,2) \ , \quad \psi^{b}_r \ \ (b=3,4)\ , \quad \hbox{with} \ \ 
-\tfrac{w-1}{2} \leq r \leq \tfrac{w-1}{2} \ ,
\ee
as well as all the modes of the form 
\be\label{outofwedge}
Z^I_r \ \hbox{and} \ (Y_J)_r  \qquad \hbox{with $r\leq - \frac{w+1}{2}$.}
\ee
(All the remaining modes annihilate $|0\rangle_w$.)
Note that for $w=1$, the modes in (\ref{wedge}) are just the zero modes of the Ramond sector representation. We also want to think of these wedge modes for general $w>0$ as generalised zero modes (or topological modes), while the out-of-the-wedge modes in (\ref{outofwedge}) are non-zero  modes.\footnote{This is somewhat reminiscent of what happened in the analysis of  ${\cal W}_\infty$ algebra representations, see e.g.\ \cite{Gaberdiel:2011zw}.} 

The intuition behind this viewpoint is that the spectrally flowed sectors describe monopole solutions  \cite{Dolan:2007vv,Nair:2007md}  of the associated twistor theory \cite{Witten:2003nn,Berkovits:2004hg}, and that these monopole solutions have holomorphic (zero) modes as described in  \cite[eq.~(2.14)]{Dolan:2007vv} and  \cite[eq.~(5)]{Nair:2007md}. We can also arrive at a similar conclusion in the ${\rm AdS}_3$ case where the covering map that characterises the $2$-point function of two $w_i$-spectrally flowed vertex operators --- this is what corresponds to the classical configuration space --- has 
\be\label{count1}
2N+2 = 4 + \sum_{i=1}^{2} (\hat{w}_i - 1) =  \sum_{i=1}^{2} w_i 
\ee
a priori undetermined coefficients, see eq.~(4.2) of \cite{Dei:2020zui}.\footnote{Since in \cite{Dei:2020zui} spectral flow is counted relative to the R-sector representation, we denote it by $\hat{w}_i$, where $\hat{w}_i = w_i-1$. The covering map is a ratio of polynomials of degree $N$, and hence has $2N+2$ free parameters. Our conventions for the wedge modes are explained in Section~\ref{sec:AdS3}.} From a microscopic perspective, these $2w$ parameters ($w=w_1=w_2$) can be associated with the positive wedge modes of $\xi^+$ and $\eta^+$ acting on either of the two vertex operators. However, using a contour deformation argument, we can also think of the positive wedge modes that act on the out-state at infinity in terms of the corresponding negative wedge modes acting on the in-state at zero. In this way, the $2w$ classical parameters can be identified with all the wedge modes acting on the in-state at zero.

\section{Physical state conditions}\label{Sec:3}

Next we need to explain which states of this worldsheet theory should describe physical string states. At this stage we do not have a detailed understanding of the relevant cohomology, so the following discussion will be somewhat sketchy. In a sense, this was also the case for our ${\rm AdS}_3$ analysis in \cite{Eberhardt:2018ouy}, although, because of the work of \cite{Berkovits:1999im}, the relevant BRST operators are in principle known in that context. The translation of that discussion into the free field worldsheet realisation, which is relevant to the tensionless limit, and the corresponding  cohomology is currently being explicitly worked out \cite{GGVit}. 

Guided by this example, we will be making a well-motivated (albeit only educated) guess for what the cohomology analysis will amount to in the present situation. In support of our proposal, we will show in Section~\ref{sec:spectrum} that it leads to the correct spectrum of free super Yang-Mills in 4d. Moreover, there is also a natural ${\rm AdS}_3$ analogue for our prescription, and we will show in Section~\ref{sec:AdS3} that this leads to (a natural subsector of) the symmetric orbifold spectrum.

\subsection{The hybrid formalism for ${\rm AdS}_3$}

In order to motivate our proposal it is instructive to first revisit the hybrid description for ${\rm AdS}_3$ \cite{Berkovits:1999im} that underlies the analysis in \cite{Eberhardt:2018ouy}. In the free field description of that theory, see in particular \cite{Dei:2020zui}, we have $4$ symplectic bosons, $4$ real (or $2$ complex) fermions, as well as a topologically twisted $\mathbb{T}^4$ theory. 
The four bosons and four fermions, which come in two canonically conjugate pairs, give a realisation of the $\mathfrak{psu}(1,1|2)_1$ WZW theory describing the tensionless limit of the ${\rm AdS}_3\times {\rm S}^3$ sector.   

The approach of \cite{Berkovits:1999im} uses the idea that a conventional RNS $N=1$ string background (with central charge $c=15$) can be equivalently viewed as a critical $N=2$ string background (with $c=6$) after including a twisted version of the $N=1$ ghosts \cite{Berkovits:1993xq}. In fact, the physical cohomology and scattering amplitudes of this theory (calculated using the usual worldsheet prescription for an $N=2$ theory) are identical to those of the original $N=1$ theory \cite{Berkovits:1993xq, Ohta:1994qp}. It was also realised in \cite{Berkovits:1994vy} that an equivalent (and simpler) way to obtain the $N=2$ string physical cohomology and scattering amplitudes is via a $N=4$ topological string theory.\footnote{This $N=4$ refers to the global symmetry of the background.} The latter theory is a generalisation of the more familiar (global) $N=2$ twisted topological string. 

The other ingredient in the approach of \cite{Berkovits:1999im} is the hybrid description \cite{Berkovits:1994wr} which (originally in flat space)  employs a field redefinition to convert (some of) the RNS matter and ghost variables of an $N=1$ critical string to Green-Schwarz variables. This makes some of the spacetime supersymmetry manifest --- in particular, eight of the maximal sixteen supersymmetries in six dimensions. Together with the embedding in the $N=2$ string described in the previous paragraph this then allows one to exhibit the spacetime supersymmetries of the ${\rm AdS}_3\times {\rm S}^3$ background. The gauged $N=2$ currents  can be written down explicitly in terms of the currents of the $\mathfrak{psu}(1,1|2)$ supergroup, as well as a couple of additional ghost fields. Furthermore, as in the RNS setup, the theory can either be viewed as a (twisted) critical $N=2$ string, or as a $N=4$ topological string with respect to an extended global $N=4$ algebra. In this latter viewpoint the physical spectrum is  given by a double cohomology of the two BRST-like operators, one arising from the current $G^+(z)$ (which is part of the $N=2$ structure), and the other from $\tilde{G}^+(z)$ (which is one of the additional generators of the global $N=4$ symmetry).  

For the level one $\mathfrak{psu}(1,1|2)_1$ WZW theory that is of interest here, the matter fields consist, as mentioned above, of $4$ symplectic bosons and $4$ real (or $2$ complex) fermions. These actually realise a 
$\mathfrak{u}(1,1|2)_1$ WZW which  needs to be quotiented out by the overall $\mathfrak{u}(1)$ in order to yield $\mathfrak{psu}(1,1|2)_1$. The symplectic bosons and fermions have central charge $c=0$ (both the bosons and fermions have conformal weight $\frac{1}{2}$). In addition we have 
\begin{itemize}
\item the conformal  $(2,-1)$ $bc$ diffeomorphism ghost system bosonised into a scalar $\sigma$ 
with $c=-26$ 
\item the $\mathfrak{u}(1)$ $(1,0)$ ghost system with $c=-2$
\item another scalar $\rho$ which is a bosonised combination of the original $(\frac{3}{2}, -\frac{1}{2})$ superdiffeomorphism ghosts. It has a central charge $c=28$, and can be viewed as the sum of two twisted $(\beta^{\pm}, \gamma^{\mp})$ systems with $c=26$ and $c=2$, respectively.
\end{itemize}
Altogether we therefore have a critical string background with net vanishing central charge. One can construct the $N=2$ currents for this free field theory \cite{GGVit}, and it can be either viewed as a (twisted) critical $N=2$ string, or as a topological $N=4$ string. The former viewpoint makes the counting of physical states quite transparent: 
 before gauge fixing there are $8$ bosonic ($4$ symplectic bosons and $4$ bosons from the $\mathbb{T}^4$) and $8$ fermionic ($4$ free fermions from the free field description of $\mathfrak{u}(1,1|2)$, and $4$ free fermions from the $\mathbb{T}^4$) degrees of freedom. Gauging the $N=2$ superconformal symmetry removes $4+4$ of these degrees freedom --- this is what the ghosts enumerated above effectively do, see \cite{Eberhardt:2018ouy} eq.~(E.6) and below. Thus we end up with the $4+4$ degrees of freedom\footnote{For completeness let us mention how the counting works for $k>1$. Now we have six bosons from the ${\rm AdS}_3\times{\rm S}^3$, the two bosons $(\rho, \sigma)$ and eight fermions. We also have the four bosons and four fermions from the $\mathbb{T}^4$. The $N=2$ gauging removes four bosons and four fermions leaving a total of eight bosons and eight fermions as might be expected of a full fledged 10d theory.}  that appear say in \cite[eq.~(5.4)]{Eberhardt:2018ouy}, and these account for the excitation modes of the symmetric orbifold of $\mathbb{T}^4$.

\subsection{Gauging an ${\cal N}=4$ superconformal symmetry}

Let us now try to generalise these ideas to the case at hand. In our worldsheet theory 
we have $8$ symplectic bosons and $8$ real fermions from the free field realisation of $\mathfrak{u}(2,2|4)_1$.\footnote{Note that now there is no analogue of a $\mathbb{T}^4$ factor, so this is all there is.} The total central charge of this matter system is again $c=0$. Since the matter content is just double that of the ${\rm AdS}_3\times {\rm S}^3$ case, we can expect a generalisation of the $N=2$ construction from above. In particular, the construction in terms of the free fields can now be applied to construct two sets of $G^+$ supercurrents for the $\mathfrak{u}(1,1|2)\oplus\mathfrak{u}(1,1|2) \subset \mathfrak{u}(2,2|4)$. 

We take the presence of additional worldsheet supercurrents as an indication that there exists a construction where the erstwhile $N=2$ structure is extended to a small $N=4$ worldsheet superconformal symmetry. In fact, there is an embedding of the $N=2$ critical string (together with its twisted ghost system) into a $N=4$ critical string \cite{Ohta:1995hw}. It is natural to expect that this would preserve the full $\mathfrak{u}(2,2|4)$ symmetry as realised on the matter fields. We would further need to take a quotient by the overall $\mathfrak{u}(1)$ as in the ${\rm AdS}_3\times {\rm S}^3$ case. 
Our proposal therefore is that there is a generalisation of the BVW construction \cite{Berkovits:1999im} with worldsheet $N=4$ gauge symmetry which makes the $\mathfrak{psu}(2,2|4)$ symmetry manifest. 

In order for this to work the twisted $N=4$ ghost system must have total central charge $c=0$, see also \cite{Baulieu:1996mr}, and this is indeed the case since the relevant ghosts are  
\begin{itemize}
\item the conformal $(2,-1)$ $bc$ ghost system with $c=-26$ 
\item the $J^{++}$  $(2,-1)$ $bc$ ghost system with $c=-26$
\item the $J^3$  $(1,0)$ $bc$ ghost system with $c=-2$
\item the $J^{--}$ $(0,1)$ $bc$ ghost system with $c=-2$
\item two superconformal  $(2,-1)$ $\beta^{+}\gamma^{-}$ ghost systems with $c=26$ each
\item two superconformal  $(1,0)$ $\beta^{-}\gamma^{+}$ ghost systems with $c=2$ each 
\end{itemize}
Note that the relevant twist is relative to the $\mathfrak{u}(1)\subset \mathfrak{su}(2)$ R-symmetry algebra of the $N=4$ symmetry, and with respect to this $\mathfrak{u}(1)$,  the $\mathfrak{su}(2)$ currents have twice the charge of the supercurrents.  

If we take this proposal seriously we would expect that the gauging will remove $8+8$ degrees of freedom. Since our worldsheet theory only has $8+8$ degrees of freedom --- the $8$ symplectic bosons and the $8$ real fermions --- this would suggest that all excitation modes are eliminated, and that the resulting theory is topological. This ties in with previous proposals, see in particular \cite{Berkovits:2019ulm}. 

More specifically, the gauging should remove all non-zero (out-of-the-wedge) modes, i.e.\ all the modes in (\ref{outofwedge}), but it is natural to believe that we retain all the zero (or wedge) modes, i.e.\ all the modes in (\ref{wedge}). Furthermore, we postulate that these modes are to be thought of as left-movers, as is, for example, manifest from the analysis in \cite{Dolan:2007vv} in the related twistor string. (In that case, it was only the left movers which had zero modes in the presence of $w>0$ units of $\mathfrak{u}(1)$ flux on the worldsheet.) In fact, there seems to be a close relation between the spectral flow in our setup, and the  $\mathfrak{u}(1)$ flux in the conventional twistor string, although the details are not clear to us: in the twistor case, the moding of the different fields is shifted according to their ${\cal C}_0$ charge, while spectral flow shifts the mode numbers  depending on their ${\cal D}_0-{\cal R}_0$ charge, see eq.~\eqref{tildeL}.  It would be interesting to understand the relation between the two twists in more detail; this should also shed light on why only modes of a specific chirality survive after spectral flow. 

In any case, we claim that, for each $w\geq 0$, the physical states are generated by the action of the (left moving) wedge modes in (\ref{wedge}) on the `vacuum' state $|0\rangle_w$. Since the wedge modes carry charge with respect to ${\cal C}_0$, not all of these descendants will be part of the $\mathfrak{psu}(2,2|4)$ theory, and we therefore still need to impose at least the ${\cal C}_0=0$ condition. However, given that in the $w$ spectrally flowed sector the set of zero modes is somewhat larger, it is natural that ${\cal C}_0=0$ is not sufficient, and that we should instead impose ${\cal C}_n=0$ for all $n\geq 0$. (Because we are only considering the wedge mode descendants, only the conditions ${\cal C}_n=0$ for $n=0,\ldots, w-1$ are non-trivial.) 
We expect these conditions to arise from the physical cohomology as residual gauge conditions on the physical wedge modes. Similarly, we also expect the original Virasoro condition, when restricted to the physical modes, to be of the form $L_0=0$ mod$(w)$, i.e.\ $L_0$ must be an integer multiple of $w$. This condition is very reminiscent of the Virasoro condition on the transverse degrees of freedom in light-cone gauge where $L_0^{\perp} \propto p^+$. The role of the light-cone momentum $p^+$ is played here by $w$, which is consistent with the pp-wave light-cone quantisation which is central to the BMN picture \cite{Berenstein:2002jq} as well. We will comment more on this in the next section --- see the discussion around eq.~\eqref{oscnum}. 

All in all, we arrive at the following proposal for the physical states of this worldsheet theory:

\begin{verse}
  
In the $w$ spectrally flowed sector the physical states are the ${\cal C}_n$ primary states in the Fock subspace that is generated by the wedge modes (\ref{wedge}) from the 
 `vacuum' state $|0\rangle_w $. In addition they have to satisfy the mass-shell condition $L_0=0$ mod$(w)$.
\end{verse}

We will show in Section~\ref{sec:spectrum} that this proposal reproduces exactly the single trace spectrum of free super Yang-Mills in 4d, where the spectral flow parameter $w$ is to be identified with the number of fundamental fields (or letters) in the trace. We have also analysed the corresponding states for the worldsheet theory describing strings on ${\rm AdS}_3 \times {\rm S}^3 \times {\mathbb M}$ at $k=1$. In that case, the above prescription leads to a fairly natural subset of physical states in the symmetric orbifold of ${\mathbb M}$ --- note that we are only considering here excitations that come from the ${\rm AdS}_3 \times {\rm S}^3$ factor, and hence we do not expect to see the entire symmetric orbifold spectrum. This analysis is described in Section~\ref{sec:AdS3} and  it thus demonstrates that the $N=2$ constraints of the  ${\rm AdS}_3 \times {\rm S}^3$ theory effectively reduce the physical modes to only the wedge modes with the residual gauge constraints. 
This therefore represents another non-trivial consistency check of our proposal and buttresses the belief that the quantisation procedure sketched here for the ${\rm AdS}_5 \times {\rm S}^5$ case captures the right physics.

\section{Explicit determination of the spectrum}\label{sec:spectrum}

In this section we describe the physical states of our gauge fixed worldsheet theory. This is to say, we shall consider the Fock space ${\cal F}_{\rm wedge}$ of states generated by the wedge modes (\ref{wedge}) from $|0\rangle_w$, and identify the states $\phi$ that satisfy 
\be\label{physical}
{\cal C}_n\, \phi = 0 \ \ (n\geq 0) \ , \qquad (L_0 + p w) \, \phi=0 \ \ (p\in\mathbb{Z}) \ . 
\ee

\subsection{Reproducing the free SYM spectrum}\label{sec:4.1}

To start with we want to show that the above physical states are in one-to-one correspondence with the spectrum of free ${\cal N}=4$ super Yang-Mills theory in the planar limit. Let us consider {\em all} the wedge modes of our worldsheet theory, i.e.\ all the modes 
\be
\hbox{$Z^I_r$ and $(Y_J)_r$ with} \quad -\tfrac{w-1}{2} \leq r \leq \tfrac{w-1}{2} \ . 
\ee
(This includes in addition to the wedge modes in (\ref{wedge}) also their conjugate modes; note that these conjugate wedge modes all annihilate $|0\rangle_w$, as follows from eq.~(\ref{prop2}).) We want to think of them as the `momentum modes' $\Phi_r$ associated to `position operators' $\hat{\Phi}_j$ in a discrete space, with the two sets of modes being related by a discrete Fourier transform
\be\label{disF}
\Phi_r = \frac{1}{\sqrt{w}}\, \sum_{j=1}^w \hat{\Phi}_j \, e^{2\pi i\frac{rj}{w}} \ , \qquad 
\hat{\Phi}_j = \frac{1}{\sqrt{w}}\, \sum_{r=- (w-1)/2}^{(w-1)/2} \Phi_r \, e^{-2\pi i\frac{rj}{w}}  \ , 
\ee
where $\hat{\Phi}_{j} = \hat{\Phi}_{j+ nw}$ for $n\in \mathbb{Z}$. Thus we may take $j$ to run from $j=1,\ldots,w$, and we can think of $j$ as labelling the different sites of a spin chain of length $w$. The only non-trivial commutators of the momentum modes have the  schematic form 
\be
{}[ Z_r^I,(Y^\dagger_J)_s]_{\pm}  = \delta^{I}_{J} \, \delta_{r,-s} \ , 
\ee
and this implies that the corresponding position operators satisfy 
\be
{}[ \hat{Z}_{j_1}^I,(\hat{Y}^\dagger_J)_{j_2}]_{\pm}  = \delta^I_J \,   \delta_{j_1,j_2}\ .
\ee
This is to say, the operators at different positions commute with one another. We also note that 
\be\label{L0}
[e^{\frac{2\pi i}{w}\, L_0}, \hat{\Phi}_j ] = \hat{\Phi}_{j+1} \ . 
\ee
Thus the condition that $L_0= -pw$ for $p\in\mathbb{Z}$ simply means that we only consider cyclically invariant combinations of position space operators along the spin chain. 

It remains to understand the meaning of the ${\cal C}_n\phi=0$ condition. Let us define, for each site  $j$, the $\hat{\cal C}_j$ generator via 
\be
\hat{\cal C}_j = \frac{1}{2} \sum_I :  (\hat{Y}_I)_j \, \hat{Z}^I_j : \ , 
\ee
where normal ordering here means that we move the (non-trivial) wedge modes (\ref{wedge}) to the left of the $|0\rangle_w$-annihilating wedge modes. By construction, these generators satisfy 
\be
[\hat{\cal C}_{j_1}  , \hat{\Phi}_{j_2}] = \pm \tfrac{1}{2}  \hat{\Phi}_{j_2} \, \delta_{j_1,j_2} \ ,
\ee
and they annihilate the ground state $|0\rangle_w$. We then consider the linear combinations
\begin{eqnarray}
\sum_{j=1}^{w} e^{2\pi i \frac{j n}{w}}\,  \hat{\cal C}_j  & = & \frac{1}{2 w} \, \sum_{j=1}^{w}  \, \sum_{r,s=-(w-1)/2}^{(w-1)/2} \, \sum_I \, 
: (Y_I)_r \, Z^I_s :  \, e^{- 2\pi i  \frac{j(r+s-n)}{w}} \\
& = & \frac{1}{2}\, \sum_{r=n-(w-1)/2}^{(w-1)/2} \, \sum_I \, : (Y_I)_r \, Z^I_{n-r} : \ ,  \label{Ccond}
\end{eqnarray}
where $n=0,1,\ldots, w-1$, and we have plugged in the definition (\ref{disF}) from above. The right-hand-side acts exactly as ${\cal C}_n$ on the wedge Fock space, and thus requiring that ${\cal C}_n=0$ for $n=0,1,\ldots, w-1$ --- the higher modes vanish automatically ---  is equivalent to demanding that each $\hat{\cal C}_j=0$. In turn this implies that at the $j$'th site we are only allowed to consider states generated by the $\hat{\Phi}_j$ with total $\hat{\cal C}_j$ charge zero. Since such states are in one-to-one correspondence with the states in the singleton representation of $\mathfrak{psu}(2,2|4)$, see Section~\ref{sec:Ramond}, this implies that the physical states are in the $w$'th tensor product of the singleton representation subject to the cyclicity constraint imposed by the $L_0$ condition. This is precisely how the single trace operators are built from the SYM letters, and thus we automatically reproduce the correct spectrum of ${\cal N}=4$ SYM in the planar limit, see e.g.\ \cite{Beisert:2004di}. 

For $w=0$, the only physical state is the NS vacuum, and it has the quantum numbers of the identity operator of the SYM theory. For $w=1$, the wedge modes are just the zero modes, and hence the physical states sit in the singleton representation, see Section~\ref{sec:Ramond}. These degrees of freedom are only present in a ${\rm U}(N)$ gauge theory, and will need to be projected out if we are restricting to the ${\rm SU}(N)$ gauge group. Below we will discuss more general $w$ and how the worldsheet description (in particular, through its $L_0$ eigenvalues) gives a novel way of organising the SYM spectrum which is a covariant generalisation of the BMN basis. 

\subsection{The DDF operators}

As a sanity check of the above argument we have also performed a direct analysis of the physical state conditions (partially using {\tt Mathematica}), and we describe our explicit checks in Appendix~\ref{app:B}. In doing so we have noticed that the resulting spectrum always has a simple form: all physical states that we have found can be obtained by the action of a certain family of DDF-like operators from the ground state. These DDF-like operators are defined via 
\be\label{Sdef}
S^{{\bf a}}_m \equiv (S_{I}{}^{J}) _m = \sum_{r =  m - \frac{w-1}{2}}^{\frac{w-1}{2}} (Y_I)_{r} \, (Z^J)_{m-r} \ , 
\ee
where $0\leq m \leq w-1$, $Y_ I= (\mu^\dagger_{\alpha},\lambda^\dagger_{\dot{\alpha}},\psi^\dagger_a)$ and $Z^I = (\lambda^{\alpha},\mu^{\dot{\alpha}},\psi^a)$ are defined as before, and ${\bf a}$ collectively denotes the indices $(I,J)$. Since $m\geq 0$, these generators map ${\cal F}_{\rm wedge}$ to itself. Furthermore, since on ${\cal F}_{\rm wedge}$ the $S^{{\bf a}} _m$ modes with $m\geq 0$ commute with ${\cal C}_n$, they map physical states to physical states, except that they modify generically (i.e.\ unless $m=0$) the mass-shell condition, $(L_0 + p w) \, \phi=0$. We mention in passing that the $S^{{\bf a}} _m$ modes satisfy the super Lie algebra
\be
[S^{{\bf a}} _m, S^{{\bf b}} _n]_{\pm} = \left\{ \begin{array}{cl}
f^{{\bf a} {\bf b}}{}_{{\bf c}} \, S^{{\bf c}}_{m+n} \quad & m+n \leq w-1 \ , \\
0 \quad & m+n \geq w \ ,
\end{array} \right.
\ee
where $f^{{\bf a} {\bf b}}{}_{{\bf c}}$ are the structure constants of $\mathfrak{u}(2,2|4)$; this is somewhat reminiscient of a Yangian algebra, except that in our case the right-hand-side of the Serre relation, see e.g.\ \cite[Appendix~A]{Bernard:1990jw}
\begin{eqnarray}
& & [S^{{\bf a}} _1, [S^{{\bf b}} _1,S^{{\bf c}} _0]] + [S^{{\bf b}} _1, [S^{{\bf c}} _1,S^{{\bf a}} _0]] + [S^{{\bf c}} _1, [S^{{\bf a}} _1,S^{{\bf b}} _0]] \\
& & \qquad \qquad = \frac{1}{24} f^{{\bf a}{\bf d}{\bf k}} \,  f^{{\bf b}{\bf e}{\bf l}} \, f^{{\bf c}{\bf f}{\bf m}} \, f_{{\bf k}{\bf l}{\bf m}} \, \{ (S_{{\bf d}}) _0,  (S_{{\bf e}}) _0, (S_{{\bf f}}) _0 \} 
\end{eqnarray}
vanishes identically. On the other hand, our algebra does satisfy a number of interesting higher order identities; for example we have --- this follows directly from the definition in eq.~(\ref{Sdef}) upon writing out both sides 
\be\label{SS}
(S_{I_1}{}^{J_1})_{w-1} \, (S_{I_2}{}^{J_2})_{w-1} = \pm (S_{I_1}{}^{J_2})_{w-1} (S_{I_2}{}^{J_1})_{w-1} \ , 
\ee
where the sign depends on the parity of the indices. There are also similar higher order identities with smaller mode numbers. 

The zero modes $S^{{\bf a}} _0$ are in fact the generators of the global $\mathfrak{psu}(2,2|4)$ algebra. Since they map physical states to physical states, they guarantee that the physical spectrum falls into $\mathfrak{psu}(2,2|4)$ representations.\footnote{Since ${\cal C}_0=0$ on physical states, we get a representation of $\mathfrak{psu}(2,2|4)$ and not just of $\mathfrak{u}(2,2|4)$.} The simplest physical state is the generating state $|0\rangle_w$ itself, which satisfies by construction, see eq.~(\ref{tildeC})
\be
{\cal C}_n \, |0\rangle_w = 0 \ \ n\geq 0 \ .
\ee
Because of (\ref{Hiflow}) and (\ref{DRspecflow}), the resulting state generates thus the highest weight representation 
\be
\bigl[ 0,0; [0,w,0] \bigr]_{w} \ ,
\ee
where we have used the same notation as in (\ref{singleton1}). We have checked that all physical states at $L_0=0$ lie in this BPS representation. We note for later that in the BMN regime, corresponding to large $w$, this state is the BMN light cone vacuum state with $p^+=w$.

\subsection{The low-lying spectrum}

For $w=2$ we have already given an explicit description of all the physical states in \cite{Gaberdiel:2021iil}; in particular, as is explained there, at $L_0=-2p$, the physical states lie in the $\mathfrak{psu}(2,2|4)$ representation, 
\be\label{hspin}
L_0=-2p : \qquad \bigl[ p-1,p-1;[0,0,0] \bigr]_{2p} \ , 
\ee
and this accounts for all the states in free SYM with $w=2$.  These states correspond to the higher spin conserved currents (twist two operators) for $p>1$. 
On the other hand, the state with $p=1$ has a scalar highest weight state, and it is to be identified with the Konishi multiplet. As was explained in \cite{Gaberdiel:2021iil}, all the states in \eqref{hspin} can be obtained by the action of the DDF-like operators $S^{{\bf a}} _n$ with $n=0,1$ from the (BPS) generating state $|0\rangle_2$. 

For larger values of $w\geq 3$, the spectrum at $L_0=-pw$ is quite complicated, and we do not have  closed formulae for it. However, we have solved the physical state conditions (\ref{physical}) explicitly for the low-lying states, and we have found that all physical states can be obtained by the action of the DDF-like operators $S^{{\bf a}} _n$ with $n=0,1,\ldots ,w-1$ from the generating state $|0\rangle_w$. It should be possible to find a counting formula for all of these states, although, because of the non-linear relations of the kind (\ref{SS}), this is not entirely straightforward. A summary of some of our explicit results can be found in  Appendix~\ref{app:B}. In particular, we have analysed the low-lying states for $w=3$,  as well as certain subclasses of states --- those made purely from  fermionic and bosonic descendants, see Appendix~\ref{app:Cferm} and \ref{app:Cbos}, respectively ---  for higher values of $w$. For example, for 
$w=3$ and $L_0=-3$ we get the representations in eqs.~(\ref{B.5}) and (\ref{B.6}). By analogy with the situation for ${\rm AdS}_3$, see eqs.~(\ref{L6m}) and (\ref{L6mp3}), one would then expect that the general formula is for $L_0=-6m$ with $m\geq 1$
\begin{align}
& \bigoplus_{r=0}^{3m} \Biggl\{ {\cal V}_{0,3m+r} \oplus \bigoplus_{k=1}^{\infty} \Bigl ({\cal V}_{2k,3m+r} \oplus {\cal V}_{-2k,3m+r}  \Bigr) \Biggl\} \nonumber \\
& \qquad \bigoplus_{r=0}^{3m-2} \Biggl\{ {\cal V}_{-1,3m+r-1} \oplus {\cal V}_{1,3m+r-1} \oplus \bigoplus_{k=1}^{\infty} \Bigl ({\cal V}_{2k+1,3m+r-1} \oplus {\cal V}_{-2k-1,3m+r-1}  \Bigr) \Biggl\} 
\end{align}
while for $L_0=-6m-3$ with $m\geq 0$ we get instead 
\begin{align}
& \bigoplus_{r=0}^{3m+1} \Biggl\{ {\cal V}_{0,3m+r+2} \oplus \bigoplus_{k=1}^{\infty} \Bigl ({\cal V}_{2k,3m+r+2} \oplus {\cal V}_{-2k,3m+r+2}  \Bigr) \Biggl\}  \nonumber \\
& \qquad \qquad \bigoplus_{r=0}^{3m} \Biggl\{ {\cal V}_{-1,3m+r} \oplus {\cal V}_{1,3m+r} \oplus \bigoplus_{k=1}^{\infty} \Bigl ({\cal V}_{2k+1,3m+r} \oplus {\cal V}_{-2k-1,3m+r}  \Bigr) \Biggl\} \ , 
\end{align}
where we have used the notation of \cite{Beisert:2004di}. Comparing with, say eq.~(3.8) of  \cite{Beisert:2004di}, we see that the worldsheet gives a refined and more natural organisation of the physical states in terms of $L_0$ eigenvalues. 

It would be interesting to understand better the systematics of this organisation of the SYM spectrum in terms of the eigenvalues of $L_0$. As remarked earlier, it appears to be a covariant version of the BMN basis in terms of the light-cone Hamiltonian (in that case, for a large number, $w$, of letters). The number of bits or letters $w$ is indeed what $p^+$ is roughly like. This also fits in with the picture that the Yang-Mills double line contractions are like light-cone strips --- each of width one and thus of width $w$ in a two point function of operators built from $w$ letters, see \cite{Gopakumar:2003ns,Gopakumar:2004qb,Gopakumar:2005fx, Razamat:2008zr}, as well as \cite{Gaberdiel:2020ycd,BGMR}.

In fact, if we compare with the BMN limit at large $w$, we can interpret the integer $p$ that appears in the $L_0=-pw$ condition with the light-cone Hamiltonian $H_{\rm l.c.} =2p^-$. To see this note that the simplest way to create a state with $L_0=-pw$ is to excite $m$ of our wedge modes, each with the highest allowed mode number $\tfrac{w-1}{2}$, so that 
\be\label{oscnum}
m\frac{w-1}{2} \approx m\frac{w}{2} = pw \ ,
\ee 
and hence $m \approx 2p$. On the other hand, the twistor modes have to appear in pairs so as to satisfy the ${\cal C}_0=0$ constraint,  and the (${\cal C}_0=0$)-neutral pairs can be identified with the BMN oscillators. Thus, to leading order in $w$, we have $N_{\rm tot}=p$ BMN oscillators. Finally, we note that in the limit where the SYM coupling vanishes, the number of BMN oscillators equals the light-cone Hamiltonian $H_{\rm l.c.} = 2 p^- = N_{\rm tot} $, see e.g.~\cite[eq.~(3.2) and (3.5)]{Berenstein:2002jq}. This also justifies why $p$ in  (\ref{physical}) should be an integer.

\section{The physical spectrum for ${\rm AdS}_3$}\label{sec:AdS3}

In this section we analyse the string spectrum for ${\rm AdS}_3\times {\rm S}^3$ using the approach of this paper. The main motivation for doing so is to see how the present proposal for quantising the theory in terms of the wedge modes, compares with the more firmly established results that are available for ${\rm AdS}_3$. We will see that it gives a consistent picture and thus indirectly supports the case for the analogous prescription for ${\rm AdS}_5$. 

Let us begin by reviewing the case of ${\rm AdS}_3$. It is generally believed that string theory on ${\rm AdS}_3 \times {\rm S}^3 \times \mathbb{T}^4$ is dual to a 2d CFT that lies on the same moduli space of CFTs that also contains the symmetric orbifold of $\mathbb{T}^4$. The natural analogue of free super Yang-Mills theory in 4d is the symmetric orbifold theory itself, and its dual string background was identified in \cite{Eberhardt:2018ouy}: it corresponds to the theory with minimal $k=1$ NSNS flux. This string theory can be described explicitly using the hybrid formalism of \cite{Berkovits:1999im}, and for pure NSNS flux the ${\rm AdS}_3\times {\rm S}^3$ factor is described by a WZW model corresponding to $\mathfrak{psu}(1,1|2)$. At level $k=1$, this WZW model has a free field realisation in terms of $4$ symplectic bosons and $4$ fermions, see Appendix~\ref{app:AdS3} for a brief summary. 

In the hybrid formalism the physical spectrum is described by a certain cohomology, and it would be interesting to work this out in detail for the above example \cite{GGVit}. However, in order to determine the resulting physical spectrum, one can also use a shortcut: for $k>1$ the hybrid description (with pure NSNS flux) and the RNS description of \cite{Maldacena:2000hw} are believed to be equivalent \cite{Gotz:2006qp,Troost:2011fd,Gaberdiel:2011vf, Gerigk:2012cq}, and this allows one to determine the `ghost contribution' of the hybrid string by comparison with the RNS approach (see Appendix E of \cite{Eberhardt:2018ouy}). Extrapolating this to $k=1$ then makes a prediction for the cohomology characterising the physical spectrum in the hybrid formalism, and it was shown in \cite{Eberhardt:2018ouy} that the resulting spectrum reproduces exactly the single particle spectrum  of the symmetric orbifold of $\mathbb{T}^4$ in the large $N$ limit. 

In the following we shall concentrate only on the ${\rm AdS}_3 \times {\rm S}^3$ factor that is described by the symplectic bosons and complex fermions. We want to determine the `physical' states in this subsector that are analogous to the ones we have proposed for ${\rm AdS}_5$. More specifically, using the conventions of \cite{Dei:2020zui}, see also Appendix~\ref{app:AdS3},  the  wedge modes for ${\rm AdS}_3 \times {\rm S}^3$ are 
\be
\xi^+_r \ , \quad \eta^+_r \ , \quad \psi^-_r \ , \quad \chi^-_r \ , \qquad -\tfrac{(w-1)}{2} \leq r \leq \tfrac{(w-1)}{2} \ ,
\ee
where we have used (\ref{AdS3sf}). They carry charge 
\be
\begin{array}{rclrcl}
[Z_n, \eta^+_r ] & = & \tfrac{1}{2} \eta^+_{r+n} \ , \qquad  & [Z_n,\chi^-_r ] & = & \tfrac{1}{2} \chi^-_{r+n} \\[4pt]
{}[Z_n, \xi^+_r ] & = & - \tfrac{1}{2} \xi^+_{r+n} \ , \qquad  & [Z_n,\psi^-_r ] & = & - \tfrac{1}{2} \psi^-_{r+n} \ ,
\end{array}
\ee
and act non-trivially on the $w$-spectrally flowed NS vaccum state, which we shall  denote by $|0 \rangle_w$ as before. We note that at least the positive wedge modes are precisely those which appear in the analysis of the free field Ward identities \cite{Dei:2020zui}, and it is therefore natural to think of them as generalised zero modes. These are also the modes which are excited when one constructs covering maps, see the discussion around eq.~(\ref{count1}) in Section~\ref{Sec:1}.

The physical state conditions are now, cf.\ eq.~(\ref{physical}) 
\be\label{physical3}
Z_n\, \phi = 0 \ \ (n\geq 0) \ , \qquad (L_0 + p w) \, \phi=0 \ \ (p\in\mathbb{Z}) \ , 
\ee
where $Z_n$, which is the analogue to ${\cal C}_n$ in the ${\rm AdS}_5$ case, is defined below eq.~(\ref{u112}). In order to identify the spectrally flowed representations we finally need to know how the various Cartan generators transform under spectral flow, see eq.~(\ref{AdS3sfC})
\be
\tilde{J}^3_0 = J^3_0 - \tfrac{w}{2}\ ,    \qquad  \tilde{K}^3_0 =  K^3_0 - \tfrac{w}{2} \ , 
\ee
where $J^3_0$ and $K^3_0$ are the Cartan generators of $\mathfrak{sl}(2,\mathds{R})$ and $\mathfrak{su}(2)$, respectively. Here $J^3_0$ is to be identified with the conformal scaling operator ($L_0$) in spacetime, and thus plays an analogous role to ${\cal D}_0$ in the $\mathfrak{psu}(2,2|4)$ case. 

Upon one unit of spectral flow, the NS vacuum representation becomes the Ramond sector representation, see eq.~(\ref{AdS3w1}), whose ground state has $J^3_0=\frac{1}{2}$ and $\mathfrak{su}(2)$ spin equal to $\frac{1}{2}$. All the Ramond sector ground states are physical --- and they are the only physical states --- and their character is 
\be\label{Z1AdS3}
Z^{(1)}(t,y) = \frac{1}{(1-t)} \, \Bigl[ t^{\frac{1}{2}} \, \chi_{\frac{1}{2}}(y) + 2 \, t \, \chi_{0}(y) \Bigr] \ , 
\ee
where $t$ and $y$ are the chemical potentials corresponding to $J^3_0$ and $K^3_0$, respectively. Furthermore, $\chi_j(y)$ is the spin $j$ character of $\mathfrak{su}(2)$. This is the analogue of the singleton representation of $\mathfrak{psu}(2,2|4)$.\footnote{We should note that, relative to the discussion in \cite{Eberhardt:2018ouy}, here only the representation with $\lambda=\frac{1}{2}$ appears. The Ramond sector representations with $\lambda\neq \frac{1}{2}$ do not arise from the NS sector vacuum upon spectral flow.}

The arguments of Section~\ref{sec:4.1} apply equally here, and the  physical states are therefore counted by the suitably symmetrised tensor powers of $Z^{(1)}(t,y)$. For example, for $w=2$ the counting function for the physical states is
\be\label{Z2sym}
Z^{(2)}(t,y)= \frac{1}{2} \Bigl(  \bigl( Z^{(1)}(t,y) \bigr)^2 + \tilde{Z}^{(1)}(t^2,y^2) \Bigr) \ , 
\ee
where 
\be\label{Z1tilde}
\tilde{Z}^{(1)}(t,y) = \frac{1}{(1-t)} \, \Bigl[ t^{\frac{1}{2}} \, \chi_{\frac{1}{2}}(y) - 2 \, t \, \chi_{0}(y) \Bigr] \ .
\ee
(Here the ground state in the doublet of $\mathfrak{su}(2)$ is `bosonic', while the two descendant states in the singlet of $\mathfrak{su}(2)$ are `fermionic'.) We have also confirmed by explicit computations that this reproduces the physical states for $w=2$. In fact, we can evaluate $Z^{(2)}(t,y)$ explicitly in terms of ${\cal N}=4$ characters, and find 
\be\label{w2complete}
Z^{(2)}(t,y) = \chi_0^{\rm (BPS)} (t,y)  + \sum_{n=1}^{\infty} \chi_{h=2n}^{\,{\cal N}=4} (t,y)  \ ,
\ee
where $\chi_0^{{\rm (BPS)}} (t,y)$ is the ${\cal N}=4$ vacuum character 
\be\label{N4vac}
\chi_0^{\rm (BPS)} (t,y)  =  \frac{1}{(1-t)} \Bigl[ t \, \chi_1(y) + 2 t^{3/2} \, \chi_{\frac{1}{2}}(y) + t^2 \Bigr]  \ , 
\ee
while for $h>0$ the character $\chi_{h}^{{\cal N}=4} (t,y)$ equals\footnote{Here we are only considering the ${\cal N}=4$ wedge algebra which equals $\mathfrak{psu}(1,1|2)$; these characters are for example obtained from eq.~(2.33) of \cite{Gaberdiel:2013vva} upon considering the limit in which the large ${\cal N}=4$ algebra contracts to the small ${\cal N}=4$ algebra.}
\be\label{hmult}
\chi_h ^{\, {\cal N}=4} (t,y) = \frac{t^h}{(1-t)} \Bigl[ 1  +  2 t^{1/2} \chi_{1/2}(y) + t (\chi_1(y) + 3) + 
2 t^{3/2} \chi_{1/2}(y) + t^2 \Bigr] \ .
\ee
Thus the physical states with $w=2$ reproduce the ${\cal N}=4$ algebra, together with a certain collection of higher spin fields. (This higher spin algebra is the ${\rm AdS}_3$ analogue of the $\mathfrak{hs}(2,2|4)$ algebra of \cite{Beisert:2004di}.) More specifically, the resulting higher spin algebra consists of the small ${\cal N}=4$ algebra, together with the even spin multiplets ${\cal R}^{(s)}$ of \cite[eq.~(2.3)]{Gaberdiel:2014cha} with $s=2,4,6,\ldots$.  A more detailed description of it is given in Appendix~\ref{app:Winf}. 

In order to avoid confusion note that $w$ here is the spectral flow relative to the NS vacuum representation; it is therefore related to the spectral flow of \cite{Eberhardt:2018ouy} (which was measured with respect to the Ramond sector, and which we denote by $\hat{w}$ in the following) via
\be\label{shift}
w = \hat{w} + 1 \ . 
\ee
Thus $w=2$ corresponds to $\hat{w}=1$, which in turn describes the untwisted sector of the symmetric orbifold. 

We should note that the spacetime states counted by (\ref{w2complete}) do {\em not} describe {\em all} the physical single particle states in the untwisted sector of the symmetric orbifold. This should not come as a surprise since we are here ignoring the torus excitations, i.e.\ we have only been considering the modes associated to ${\rm AdS}_3 \times {\rm S}^3$. We therefore expect that our analysis here will only reproduce (some of) the `compactification-independent' states, i.e.\ those that do not rely on the microscopic structure of the $\mathbb{T}^4$ theory, but only use its ${\cal N}=4$ symmetry. This fits together with the fact that the higher spin algebra which we get from the $w=2$ sector is a subalgebra of that appearing in the case of $\mathbb{T}^4$ \cite{Gaberdiel:2014cha}, as well as for K3 \cite{Baggio:2015jxa}. 

In order to understand the structure of the physical states that appear for $w\geq 3$ we note that the twisted sector ground state of the $2$-cycle twisted sector of the symmetric orbifold has $h=j=\frac{1}{2}$. In particular, it therefore defines a  $\mathfrak{psu}(1,1|2)$ representation with character $Z^{(1)}(t,y)$, just as for the usual minimal (or singleton) representation generated from a free fermion, see the discussion in Appendix~\ref{app:C.4}. 

As for the ${\rm AdS}_5$ case, the physical states satisfying (\ref{physical3}) for a given $w$ are the cyclically invariant states in the $w$-fold tensor product of this minimal representation. For example, for $w=3$ these states can thus be identified with a subset of  states in the symmetric orbifold, of the form 
\be
\bar\psi^1_{-r} \psi^1_{-s} |0\rangle^{(2)} \ ,   \quad \bar\psi^1_{-r} \bar\psi^2_{-s} |0\rangle^{(2)} \ , \quad \hbox{etc.} \ ,
\ee
where the first mode runs through all the possibilities of (\ref{fund1}), while the second one is any one of the modes appearing in (\ref{fund2}), see Appendix~\ref{app:Winf} and \ref{app:C.4} for our conventions. In addition, one has to impose `cyclic symmetry'; for example, the lowest surviving state is then the upper BPS state with $h=j=\frac{3}{2}$, 
\be
\bar\psi^1_{-\frac{1}{2}} \bar\psi^2_{-\frac{1}{2}} |0\rangle^{(2)}  = J^+_{-1} \, |0\rangle^{(2)} \ . 
\ee

The analysis for $w\geq 4$ is similar. If we denote the minimal representation of $\mathfrak{psu}(1,1|2)$ by ${\cal H}_{\rm min}$, our worldsheet analysis leads to the cyclically invariant states in the tensor product 
\be\label{tensorw}
\Bigl( {\cal H}_{\rm min} \Bigr)^{\otimes w} \cong \Bigl( {\cal H}_{\rm min} \Bigr)^{\otimes (w-2)} \otimes {\cal H}_{\rm min}  \otimes {\cal H}_{\rm min}  \ . 
\ee
Let us identify the ${\cal H}_{\rm min}$ in the bracket on the right-hand-side with the $\mathfrak{psu}(1,1|2)$ descendants in the $2$-cycle twisted sector, see (\ref{halfdes}) etc., while the last two copies are identified again with the modes in (\ref{fund1}) and (\ref{fund2}), respectively. The reason for writing the left-hand-side in this manner is that the $(w-2)$-fold fusion of the $2$-cycle twisted sector contains, in particular, the $(w-1)$-cycle twisted sector, and with respect to the $\mathfrak{psu}(1,1|2)$ algebra fusion behaves essentially as a tensor product.\footnote{From the tensor product perspective, the charges are additive, and hence we only see the contribution from the $(w-1)$-cycle twisted sector. The fact that fusion behaves essentially as a tensor product with respect to the wedge algebra is familiar from e.g.\ \cite{Gaberdiel:2011zw}.} 
For example, the leading term in this $(w-2)$-fold tensor product has charges $h=j=\frac{w-2}{2}$, and hence describes the `lower' BPS state in the $\hat{w}=(w-1)$-cycle twisted sector. (Here we use that each $\hat{w}$-cycle twisted sector  in the symmetric orbifold has two BPS states: the `lower' BPS state with $h=j=\frac{\hat{w}-1}{2}$, and the `upper' BPS state with $h=j=\frac{\hat{w}+1}{2}$, see e.g.\  \cite{David:2002wn}.) Putting in the contributions from the last two factors in (\ref{tensorw}) the leading term of the entire tensor product has $h=j=\frac{w}{2}$, and hence can be identified with 
\be
\bar\psi^1_{-\frac{1}{2}} \bar\psi^2_{-\frac{1}{2}} |{\rm BPS}_{\rm lower}\rangle^{(w-1)} \ ,
\ee
which describes the `upper' BPS state in the $(w-1)$ cycle twisted sector. The other contributions from (\ref{tensorw}) then describe certain descendants, and we have schematically 
\be\label{tensorw1}
\Bigl( {\cal H}_{\rm min} \Bigr)^{\otimes w} \cong \underbrace{\Bigl( {\cal H}_{\rm min} \Bigr)^{\otimes (w-2)}}_{\in (w-1)-{\rm cycle}\ {\rm  tw.}} \otimes \ \underbrace{{\cal H}_{\rm min}  \otimes {\cal H}_{\rm min} }_{ \bar\psi^1_{-\frac{1}{2}} \bar\psi^2_{-\frac{1}{2}},\, \ldots} \ . 
\ee
The fact that the `upper' BPS state appears in this analysis is quite natural since it is this state that is to be identified with the zero'th cohomology of the 4d $\mathbb{M}$ manifold in ${\rm AdS}_3 \times {\rm S}^3 \times \mathbb{M}$, see \cite{David:2002wn}; thus this is the natural `compactification independent' state. 
\medskip

For small values of $w$ we can also be more explicit. In particular, as in \cite{Gaberdiel:2021iil} we can give a very explicit description of the physical $w=2$ spectrum. For $L_0=0$ we just have the ground state $|0\rangle_w$, together with its $\mathfrak{psu}(1,1|2)$ descendants; this gives rise to the ${\cal N}=4$ algebra, i.e.\ to the states counted by eq.~(\ref{N4vac}). For $L_0=-2p$ with $p>0$, the physical states lie in the $\mathfrak{psu}(1,1|2)$ representation generated from 
\be
\bigl(\xi^+_{\frac{1}{2}} \eta^+_{\frac{1}{2}} \bigr)^{2p -1} \, \psi^-_{\frac{1}{2}} \, \chi^-_{\frac{1}{2}} |0\rangle_2 : \qquad 
J^3_0 = 2p \ , \qquad K^3_0 = 0 \ . 
\ee
The corresponding states are therefore precisely counted by 
\be
L_0 = - 2p: \qquad \chi_{h=2p}^{{\cal N}=4} (t,y) \ .
\ee
This then accounts for all the states in (\ref{w2complete}). The corresponding analysis for $w=3$ is described in Appendix~\ref{app:AdS3w3}.

Finally, we note that the spin-chain like cyclic symmetry of the worldsheet description acts in a rather unusual manner from the viewpoint of the symmetric orbifold, see eqs.~(\ref{tensorw}) and (\ref{tensorw1}). This is probably the underlying reason why it has proven so difficult to relate the spin chain pictures that have been proposed for ${\rm AdS}_3$, see e.g.\ \cite{Sfondrini:2014via} for a review, to the symmetric orbifold.

\section{Concluding Remarks}\label{sec:concl}

Our proposal for a worldsheet description of the tensionless string on ${\rm AdS}_5\times {\rm S}^5$ has, at every stage, been informed by the much more detailed experience with the corresponding theory on ${\rm AdS}_3\times {\rm S}^3$. Our twistorial sigma model is the appropriate generalisation from $\mathfrak{psu}(1,1|2)$ to $\mathfrak{psu}(2,2|4)$, as is the spectral flow. In both cases, there is a natural division of the oscillators into `wedge' and `out-of-the-wedge' modes. 

In this paper, building on \cite{Gaberdiel:2021iil}, we proposed a way of quantising the theory on 
${\rm AdS}_5\times {\rm S}^5$ where the physical degrees of freedom arise purely from the wedge (i.e.\ generalised zero) modes. While this was not how we had originally viewed the corresponding string theory on  ${\rm AdS}_3\times {\rm S}^3$, the topological nature of the latter theory can be attributed to only the wedge modes being excited and thus motivates such a 
proposal for ${\rm AdS}_5$.  
In fact, in the previous section, we revisited the ${\rm AdS}_3$ spectrum and studied the quantisation of the wedge modes in that case, subject to the analogous residual constraints. As we explained, this gives the part of the symmetric orbifold spectrum that one might expect to be independent of the extra compactification from the 10d theory to ${\rm AdS}_3\times {\rm S}^3$, thus further cementing the intuition about the wedge modes. 

Even with all the direct and indirect evidence in support of the wedge oscillators of the twistorial sigma model being the physical modes, it would, of course, be very desirable and important to obtain this from first principles. Again, we have guidance from ${\rm AdS}_3\times {\rm S}^3$ where the sigma model is essentially quantised as an $N=2$ critical string. As discussed in Section~\ref{Sec:3}, it is natural to expect a generalisation of this quantisation scheme to our ${\rm AdS}_5\times {\rm S}^5$ sigma model, which would come with a correspondingly enhanced set of worldsheet constraints. We expect to be able to flesh this picture out in the near future \cite{GGVit}. 

Let us indicate some of the other questions that deserve a better understanding. One set of questions has to do with the stringy organisation of the SYM spectrum that the wedge oscillator construction suggests. As remarked, this gives a $\mathfrak{psu}(2,2|4)$ covariant version of the BMN picture which is not tied to a large R-charge or light-cone like limit. More specifically, the worldsheet perspective leads to a BMN-like basis in terms of momentum eigenstates built from cyclically symmetrical bits. This organisation in terms of $L_0$ eigenvalues may well turn out to be a useful way to stratify the spectrum which will generalise to nonzero coupling. 
This is what the integrable spin chain picture for the dual SYM theory also indicates \cite{Beisert:2004ry, Beisert:2010jr}. In understanding this organisation of the spectrum better, the DDF operators defined in eq.~(\ref{Sdef}) are likely to play an important role, especially in generalising to 
nonzero coupling. However, we need to understand the precise way in which they generate the spectrum, given the large number of relations they obey. 

Another property of the free theory which we have only tangentially touched upon is the higher spin $\mathfrak{hs}(2,2|4)$ symmetry which also organises the free SYM spectrum \cite{Beisert:2004di}. We have already seen the generators of this symmetry in the twist two operators of the $w=2$ sector. In fact, the worldsheet realisation gives a natural oscillator construction of $\mathfrak{hs}(2,2|4)$. This could be the starting point for understanding better the worldsheet manifestations of the spacetime higher spin symmetry. 
It will also be interesting to study the analogous questions in ${\rm AdS}_3\times {\rm S}^3$ where there are additional new features to the higher spin symmetry including their enhancement to ${\cal W}_{\infty}$ symmetries \cite{Henneaux:2010xg, Campoleoni:2010zq} and the presence of an even bigger structure (at least for the ${\mathbb T}^4$ case) ---  the higher spin square \cite{Gaberdiel:2015mra, Gaberdiel:2015wpo, Raeymaekers:2016mmm}. 

While we have concentrated on the spectrum, it is natural to ask how these considerations generalise to correlation functions. In the twistorial ${\rm AdS}_3\times {\rm S}^3$ sigma model, it was seen from the Ward identities that correlators localise to the points in moduli space where holomorphic covering maps exist. Furthermore, the resulting covering maps could be read off directly from the twistor fields \cite{Dei:2020zui}. A similar analysis should be possible for the ${\rm AdS}_5\times {\rm S}^5$ sigma model, and this is currently being pursued \cite{GGKM}. Since many of the broad features are expected to go through similarly, we ought to see an analogous localisation. From a complementary analysis of classical solutions in the ${\rm AdS}_5\times {\rm S}^5$ twistor space one sees how the notion of covering maps generalises to this case \cite{BGMR}. The resulting picture appears to tie in nicely with the 
free SYM Feynman diagram contributions to correlators: each distinct Wick contraction of the latter corresponds to an individual twistor covering map, each 't~Hooft double line is a single sheet of the covering, and the closed string worldsheet glues these rigid individual coverings together. This is again the analogue of what was already seen for the correlators of the symmetric product CFT, at least in the limit of  large $w_i$, in \cite{Gaberdiel:2020ycd}. In that case, one had a precise realisation of open-closed string duality through a (discrete) Strebel parametrisation of moduli space by individual Feynman diagrams \cite{Gopakumar:2003ns, Gopakumar:2004qb, Gopakumar:2005fx, Razamat:2008zr} and we expect this to work similarly here as well.   

Once one turns on interactions in the SYM theory, the picture will be more involved. It is likely that, just as for the spectrum, we will make contact with the integrability approach to correlators. In particular, it is very interesting to note that the hexagon approach to SYM correlators \cite{Basso:2015zoa} is most naturally formulated in terms of bilinears of $\mathfrak{su}(2|2)$ bits --- these seem to be closely related to our covariant twistorial wedge modes. In our case we have the worldsheet $\mathfrak{u}(1)$ gauge field which enforces singlets of these twistor bits and thus glues the two copies of the hexagon together. It will be very good to make the connection more precise.\footnote{We thank Pedro Vieira for pointing out the similarity of the twistor bits to what appears in the hexagon construction.} Another useful case to consider, intermediate in complexity when we turn on couplings, might be the fishnet theories which are solvable and where one has control over the number of Feynman diagrams that contribute, see e.g.\ \cite{Gromov:2019jfh}. 

Another set of questions have to do with the precise relation to the usual twistor string theories \cite{Witten:2003nn, Berkovits:2004hg} that describe perturbative SYM scattering amplitudes. For example, it is very tempting to interpret the Berkovits twistor string theory \cite{Berkovits:2004hg} as an open string version of the present theory --- the theory living on D-branes in the `zero radius' ${\rm AdS}_5\times {\rm S}^5$ string theory. This would then give a realisation of the open-closed-open string triality, advocated in \cite{joburg-talk}, in being an `F-type' open string description graph dual to the usual `V-type' SYM description. In support of a relation between the twistor open string theory and our closed string theory are the correspondences between the sector with $w$ units of $\mathfrak{u}(1)$ flux in that theory, and our $w$-spectrally flowed sector, as well as in the appearance of generalised zero modes in both cases which are again very analogous. This could potentially shed light on the fact that we have only left movers in our theory. The latter feature also seems to be similar to what one has in the related ambitwistor string constructions of \cite{Mason:2013sva}. 

The fact that it is only the set of (generalised) zero modes of the left movers that contribute is also very reminiscent of the A-model topological string. This is reflected too in the features of localisation on  moduli space for correlators that we have seen in the case of  ${\rm AdS}_3\times {\rm S}^3$ and which we expect here as well. It will be very clarifying if the twistorial sigma models can indeed be recast in terms of the more familiar topological sigma models \cite{Witten:1988xj}. This may facilitate a path integral analysis that could make the localisation onto the wedge modes more apparent, in contrast to the Hamiltonian analysis outlined in Section~\ref{Sec:3}.    

Finally, we find it tantalising that we might be close to obtaining an answer to the question raised in the introduction about the dual to free ordinary Yang-Mills theory in 4d. Pure Yang-Mills theory has a planar spectrum which is given, almost by construction, by the bosonic truncation of the wedge twistorial oscillators. Is there a version of our twistorial sigma model in which the physical degrees of freedom on the worldsheet are purely bosonic, i.e.\ only the four pairs of symplectic bosons? We find the fact that bosonic string vacua can be realised as special cases of the more supersymmetric worldsheet theories (rather than the other way around) \cite{Berkovits:1993xq} very suggestive. We need to find an appropriate way to `Higgs' the worldsheet superconformal symmetries, so as to reduce the theory to the  bosonic sector. This might well be doable.

\acknowledgments

We thank Nathan Berkovits, Andrea Dei, Lorenz Eberhardt, Abhijit Gadde, Bob Knighton, Pronobesh Maity, Juan Maldacena, Gautam Mandal, Shiraz Minwalla, Kiarash Naderi, Onkar Parrikar, Vit Sriprachyakul, Sandip Trivedi, Pedro Vieira and Spenta Wadia for useful discussions and correspondences. The work of MRG was supported by the Swiss National Science Foundation through a personal grant and via the NCCR SwissMAP. The work of RG is supported in part by the J.C.~Bose Fellowship of the DST-SERB.
RG acknowledges the support of the Department of Atomic Energy, Government
of India, under project no.~RTI4001, as well as the framework of support for the basic sciences by the people of India. 

\appendix

\section{The $\mathfrak{psu}(2,2|4)$ current algebra}\label{app:psu}

In this appendix, we give explicit expressions for the generators of  $\mathfrak{psu}(2,2|4)_1$ in terms of the free fields defined in Section~\ref{Sec:1}. We also exhibit their commutation relations, see Section~\ref{app:commur}. Finally we spell out the details of the action of the zero modes in the Ramond representation. 

\subsection{Generators of  $\mathfrak{psu}(2,2|4)_1$}
 
We begin with the bosonic subalgebra $\mathfrak{su}(2) \oplus \mathfrak{su}(2)\oplus \mathfrak{su}(4)$
\begin{eqnarray}
{\cal L}^\alpha{}_\beta & = &  \mu^\dagger_\beta \, \lambda^\alpha - \tfrac{1}{2} \delta^\alpha_\beta  \, U \\
\dot{\cal L}^{\dot{\alpha}}{}_{\dot{\beta}} & = &  \lambda^\dagger_{\dot{\beta}} \, \mu^{\dot{\alpha}} - \tfrac{1}{2} \delta^{\dot{\alpha}}_{\dot{\beta}}  \, \dot{U} \\
{\cal R}^a{}_b & = & \psi^\dagger_b \psi^a - \tfrac{1}{4} \delta^a_b \, V \ , 
\end{eqnarray}
where normal ordering is always understood, and we have introduced the generators (mimicking what we did in \cite{Eberhardt:2018ouy})
\be\label{UVdef}
U  =  \mu^\dagger_\gamma \, \lambda^\gamma \ , \qquad 
\dot{U}  =  \lambda^\dagger_{\dot{\gamma}} \, \mu^{\dot{\gamma}} \ , \qquad 
V  =  \psi^\dagger_c\psi^c \ ,
\ee
which commute with ${\cal L}^\alpha{}_\beta$, $\dot{\cal L}^{\dot{\alpha}}{}_{\dot{\beta}}$ and ${\cal R}^a{}_b$. For the following it will be convenient to introduce the combinations 
\be\label{BCD}
{\cal B}_n  =  \tfrac{1}{2} \, \bigl( U_n + \dot{U}_n \bigr) \ , \qquad 
{\cal C}_n  =   \tfrac{1}{2} \, \bigl( U_n + \dot{U}_n \bigr)  + \tfrac{1}{2} V_n  \ , \qquad 
{\cal D}_n  =  \tfrac{1}{2} \, \bigl( U_n - \dot{U}_n \bigr)   \ .
\ee
The `off-diagonal' generators of $\mathfrak{u}(2,2|4)_1$ are 
\begin{equation}\label{offdiagonal}
\begin{array}{rclrcl}
{\cal Q}^a{}_\alpha & = & \psi^a\, \mu^\dagger_\alpha   \ , \qquad \qquad & {\cal S}^\alpha{}_a & = & \lambda^\alpha\, \psi^\dagger_a\,  \ , \\[4pt] 
\dot{\cal Q}^{\dot{\alpha}}{}_{a} & = & \mu^{\dot{\alpha}} \, \psi^\dagger_a \ , \qquad \qquad & \dot{\cal S}^a{}_{\dot{\alpha}} & = &  \psi^a \, \lambda^\dagger_{\dot{\alpha}} \ ,\\[4pt] 
{\cal P}^{\dot{\alpha}}{}_{\beta} & = &  \mu^{\dot{\alpha}}\, \mu^\dagger_\beta  \ , \qquad \qquad & {\cal K}^{\alpha}{}_{\dot{\beta}} & = & \lambda^\alpha \, \lambda^\dagger_{\dot{\beta}} \ .
\end{array}
\end{equation}
We summarise the various commutation and anti-commutation relations in the following subsection. For the moment we only note 
that the generator ${\cal D}_n$ in (\ref{BCD}) is the dilatation operator of $\mathfrak{psu}(2,2|4)_1$, and its zero mode will be identified with the conformal dimension in the 4d SYM theory. On the other hand, ${\cal B}_n$ and ${\cal C}_n$ play analogous roles to $Y_n$ and $Z_n$ in \cite{Eberhardt:2018ouy,Dei:2020zui}. In particular, ${\cal C}_n$ is central (except for the non-trivial commutator with ${\cal B}_n$, see eq.~(\ref{BC}) below), and needs to be `removed' in order to go to  $\mathfrak{psu}(2,2|4)_1$ as also mentioned earlier. Following \cite{Dei:2020zui} we implement this constraint by working with the subspace on which ${\cal C}_n$ with $n\geq 0$ annihilates states; the ${\cal C}_{-n}$ descendants are then null, and are thus naturally modded out. 

As regards the $\mathfrak{su}(4)$ algebra, we use the convention that the generators of the form $({\cal R}^{a}{}_b)_0$ with $a<b$ are the positive roots, and define the Cartan generators of $\mathfrak{su}(4)$ to be
\begin{equation}\label{cartan}
H_1  =  ({\cal R}^2{}_2)_0  -  ({\cal R}^1{}_1)_0\ , \qquad 
H_2  =  ({\cal R}^3{}_3)_0  -  ({\cal R}^2{}_2)_0\ , \qquad
H_3  =  ({\cal R}^4{}_4)_0  -  ({\cal R}^3{}_3)_0 \ .
\end{equation}

\subsection{Commutation relations}\label{app:commur}

Here we will spell out the explicit commutation relations of the $\mathfrak{u}(2,2|4)_1$ generators. The bosonic generators satisfy the commutation relations
\be
{}[ ({\cal L}^\alpha{}_\beta)_m,({\cal L}^\gamma{}_\delta)_n]  =  \delta^\alpha_\delta \, ({\cal L}^\gamma{}_\beta)_{m+n} - \delta^\gamma_\beta \, ({\cal L}^\alpha{}_\delta)_{m+n} + m\, \Bigl( - \delta^\gamma_\beta \delta^\alpha_\delta + \tfrac{1}{2} \delta^\alpha_\beta \delta^\gamma_\delta \Bigr) \delta_{m,-n}  \ , 
\ee
and likewise for $(\dot{\cal L}^{\dot{\alpha}}{}_{\dot{\beta}})_m$. With the identification
\be
J^+_m = ({\cal L}^1{}_2)_m \ , \qquad J^-_m =  ({\cal L}^2{}_1)_m \ , \qquad 
J^3_m = \frac{1}{2} \Bigl( ({\cal L}^2{}_2)_m - ({\cal L}^1{}_1)_m \Bigr)\ , 
\ee
they then satisfy the commutation relations of $\mathfrak{su}(2)_{-1}$, i.e.\ 
\begin{eqnarray}
{}[J^3_m,J^3_n] & = & + \tfrac{1}{2} \, k\, m\, \delta_{m,-n} \\
{}[J^3_m,J^\pm_n] & = & \pm J^\pm_{m+n} \\
{}[J^+_m,J^-_n] & = & + 2 \, J^3_{m+n} + k \, m\, \delta_{m,-n} 
\end{eqnarray}
with $k=-1$. The same also applies to the $(\dot{\cal L}^{\dot{\alpha}}{}_{\dot{\beta}})_m$ generators, which therefore lead to another copy of $\mathfrak{su}(2)_{-1}$ with generators $\hat{J}^a_n$. On the other hand, for the $({\cal R}^a{}_b)_m$ generators we find 
\be
{}[({\cal R}^a{}_b)_m,({\cal R}^c{}_d)_m] = \delta^a_d \, ({\cal R}^c{}_b)_{m+n} - \delta^c_b \, ({\cal R}^a{}_d)_{m+n} + m\, \Bigl(  \delta^a_d \delta^c_b - \tfrac{1}{4} \delta^a_b \delta^c_d \Bigr) \delta_{m,-n}  \ ,
\ee
which are the commutation relations of $\mathfrak{su}(4)_1$. Finally, the $\mathfrak{u}(1)$ currents satisfy $[{\cal D}_m,{\cal B}_n] = [{\cal D}_m,{\cal C}_n] = 0$, as well as 
\begin{eqnarray}
{}[{\cal B}_m,{\cal B}_n] & = & - m\,  \delta_{m,-n} \\
{}[{\cal B}_m,{\cal C}_n] & = & - m \, \delta_{m,-n}  \label{BC} \\ 
{}[{\cal C}_m,{\cal C}_n] & = & 0  \\
{}[{\cal D}_m,{\cal D}_n] & = & - m\,  \delta_{m,-n} \ ,
\end{eqnarray}
since
\be
{}[U_m,U_n] = [\dot{U}_m,\dot{U}_n]  =  - 2 m \delta_{m,-n} \ , \qquad 
{}[V_m,V_n]  =  4 m \delta_{m,-n} \ . 
\ee

The `off-diagonal' generators of eq.~(\ref{offdiagonal}) transform in the manner indicated by their index structure under $\mathfrak{su}(2) \oplus \mathfrak{su}(2) \oplus \mathfrak{su}(4)$. They commute with ${\cal C}_m$,
\begin{eqnarray}
{}[{\cal C}_m,  ({\cal Q}^a{}_\alpha)_n] & = &  [{\cal C}_m,  ({\cal S}^\alpha{}_a)_n] = [{\cal C}_m,  (\dot{\cal Q}_{\dot{\alpha}\, a})_n]  = [{\cal C}_m,  (\dot{\cal S}^{\dot{\alpha}\, a})_n]  \\
& = & [{\cal C}_m,  ({\cal P}_{\alpha\dot{\beta}})_n] = [{\cal C}_m, ({\cal K}^{\alpha\dot{\beta}})_n] = 0 \ , 
\end{eqnarray}
while with respect to the other two generators we have 
\be
\begin{array}{rclrcl}
{}[{\cal B}_m,  ({\cal Q}^a{}_\alpha)_n] & = & \tfrac{1}{2} \, ({\cal Q}^a{}_\alpha)_{m+n} \ , \qquad 
& [{\cal B}_m,  ({\cal S}^\alpha{}_a)_n] & = & - \tfrac{1}{2} \, ({\cal S}^\alpha{}_a)_{m+n} \ , \\
{}[{\cal B}_m,  (\dot{\cal Q}_{\dot{\alpha}\, a})_n] & = & -  \tfrac{1}{2} \, (\dot{\cal Q}_{\dot{\alpha}\, a})_{m+n} \ , \qquad 
& [{\cal B}_m,  (\dot{\cal S}^{\dot{\alpha}\, a})_n] & = &  \tfrac{1}{2} \, (\dot{\cal S}^{\dot{\alpha}\, a})_{m+n} \ , \\
{}[{\cal B}_m,  ({\cal P}_{\alpha\dot{\beta}})_n] & = & 0 \ , \qquad 
& [{\cal B}_m, ({\cal K}^{\alpha\dot{\beta}})_n] & = & 0 \ , 
\end{array}
\ee
and
\be\label{Deigen}
\begin{array}{rclrcl}
{}[{\cal D}_m,  ({\cal Q}^a{}_\alpha)_n] & = & \tfrac{1}{2} \, ({\cal Q}^a{}_\alpha)_{m+n} \ , \qquad 
& [{\cal D}_m,  ({\cal S}^\alpha{}_a)_n] & = & - \tfrac{1}{2} \, ({\cal S}^\alpha{}_a)_{m+n} \ , \\
{}[{\cal D}_m,  (\dot{\cal Q}_{\dot{\alpha}\, a})_n] & = &  \tfrac{1}{2} \, (\dot{\cal Q}_{\dot{\alpha}\, a})_{m+n} \ , \qquad 
& [{\cal D}_m,  (\dot{\cal S}^{\dot{\alpha}\, a})_n] & = &  - \tfrac{1}{2} \, (\dot{\cal S}^{\dot{\alpha}\, a})_{m+n} \ , \\
{}[{\cal D}_m,  ({\cal P}_{\alpha\dot{\beta}})_n] & = & ({\cal P}_{\alpha\dot{\beta}})_{m+n}  \ , \qquad 
& [{\cal D}_m, ({\cal K}^{\alpha\dot{\beta}})_n] & = & - ({\cal K}^{\alpha\dot{\beta}})_{m+n}  \ . 
\end{array}
\ee
Finally, the commutation (anti-commutation) relations of the `off-diagonal' generators are 
\begin{eqnarray}
{}[({\cal K}^{\alpha}{}_{\dot{\beta}})_m,({\cal P}^{\dot{\gamma}}{}_{\delta})_n] & = & 
- \delta^{\dot{\gamma}}_{\dot{\beta}} \, ({\cal L}^{\alpha}{}_{\delta})_{m+n} + \delta^\alpha_\delta \, (\dot{\cal L}^{\dot{\gamma}}{}_{\dot{\beta}})_{m+n} 
- \delta^\alpha_\delta\, \delta^{\dot{\gamma}}_{\dot{\beta}} \, \bigl( {\cal D}_{m+n} + m \,\delta_{m,-n} \bigr) \ , \qquad \label{KP} \\ 
\{ ({\cal S}^{\alpha}{}_{a})_m,({\cal Q}^b{}_{\beta})_n \} & = & 
\delta^b_a \, ({\cal L}^{\alpha}{}_{\beta})_{m+n} + \delta^\alpha_\beta\, ({\cal R}^b{}_a)_{m+n} \nonumber \\
& & + \tfrac{1}{2} \delta^b_a\, \delta^\alpha_\beta\, \bigl({\cal D}_{m+n} + {\cal C}_{m+n} + 2m\, \delta_{m,-n} \bigr) \ ,  \\
\{ (\dot{\cal S}^{a}{}_{\dot{\alpha}})_m,(\dot{\cal Q}^{\dot{\beta}}{}_{b})_n \} & = & 
\delta^a_b \, (\dot{\cal L}^{\dot{\beta}}{}_{\dot{\alpha}})_{m+n} + \delta^{\dot{\beta}}_{\dot{\alpha}}\, ({\cal R}^a{}_b)_{m+n}  \nonumber \\
& & - \tfrac{1}{2} \delta^a_b\, \delta^{\dot{\alpha}}_{\dot{\beta}}\, \bigl ({\cal D}_{m+n} - {\cal C}_{m+n} + 2m\, \delta_{m,-n} \bigr) \ .
\label{dotSdotQ}
\end{eqnarray}

\subsection{The Ramond sector representation}\label{app:R}

In the Ramond sector the symplectic boson zero modes act on the states labelled by 
$|m_1,m_2;\dot{m}_1,\dot{m}_2\rangle$ as 
\begin{subequations}\label{laaction}
\begin{align}
\lambda^1_0 \, |m_1,m_2;\dot{m}_1,\dot{m}_2\rangle & =  2 \, m_1 \, |m_1-\tfrac{1}{2},m_2;\dot{m}_1,\dot{m}_2\rangle  \label{eq:laaction a}\\ 
\lambda^2_0 \, |m_1,m_2;\dot{m}_1,\dot{m}_2\rangle & =  2 \, m_2 \, |m_1,m_2-\tfrac{1}{2};\dot{m}_1,\dot{m}_2\rangle  \label{eq:laaction b} \\ 
(\mu^\dagger_1)_0 \, |m_1,m_2;\dot{m}_1,\dot{m}_2\rangle & =    |m_1+\tfrac{1}{2},m_2;\dot{m}_1,\dot{m}_2\rangle  \label{eq:laaction c} \\ 
(\mu^\dagger_2)_0 \, |m_1,m_2;\dot{m}_1,\dot{m}_2\rangle & =    |m_1,m_2+\tfrac{1}{2};\dot{m}_1,\dot{m}_2\rangle \ ,  \label{eq:laaction d}
\end{align}
\end{subequations}
while for the $(\mu,\lambda^\dagger)$ modes we set
\begin{subequations}\label{muaction}
\begin{align}
\mu^1_0 \, |m_1,m_2;\dot{m}_1,\dot{m}_2\rangle & =   |m_1,m_2;\dot{m}_1+\tfrac{1}{2},\dot{m}_2\rangle  \label{eq:muaction a}\\ 
\mu^2_0 \, |m_1,m_2;\dot{m}_1,\dot{m}_2\rangle & =   |m_1,m_2;\dot{m}_1,\dot{m}_2+\tfrac{1}{2}\rangle  \label{eq:muaction b} \\ 
(\lambda^\dagger_1)_0 \, |m_1,m_2;\dot{m}_1,\dot{m}_2\rangle & = - 2 \, \dot{m}_1\,  |m_1,m_2;\dot{m}_1-\tfrac{1}{2},\dot{m}_2\rangle  \label{eq:muaction c} \\ 
(\lambda^\dagger_2)_0 \, |m_1,m_2;\dot{m}_1,\dot{m}_2\rangle & =  - 2 \, \dot{m}_2 \,  |m_1,m_2;\dot{m}_1,\dot{m}_2-\tfrac{1}{2}\rangle \ .  \label{eq:muaction d}
\end{align}
\end{subequations}
For the action of the fermionic zero modes we define 
\be\label{fermizero}
\psi^a_0 \,  |m_1,m_2;\dot{m}_1,\dot{m}_2\rangle = 0 \ , \qquad a=1,2,3,4 \ . 
\ee
The eigenvalues of the $\mathfrak{u}(2,2|4)$ (Cartan) generators can then be computed from 
(\ref{laaction}) and  (\ref{muaction}). One finds that\footnote{For $U_0$ and $\dot{U}_0$ this is a matter of convention.} 
\begin{eqnarray}
J^3_0 \, |m_1,m_2;\dot{m}_1,\dot{m}_2\rangle & =  & (m_2-m_1) \, |m_1,m_2;\dot{m}_1,\dot{m}_2\rangle  \label{J3eig}\\ 
U_0 \, |m_1,m_2;\dot{m}_1,\dot{m}_2\rangle & =  & 2 \, (m_1+m_2 +\tfrac{1}{2} ) \, |m_1,m_2;\dot{m}_1,\dot{m}_2\rangle \\ 
\hat{J}^3_0 \, |m_1,m_2;\dot{m}_1,\dot{m}_2\rangle & =  & (\dot{m}_1-\dot{m}_2) \, |m_1,m_2;\dot{m}_1,\dot{m}_2\rangle  \label{J3heig} \\
\dot{U}_0 \, |m_1,m_2;\dot{m}_1,\dot{m}_2\rangle & =  & - 2 \, (\dot{m}_1+\dot{m}_2+\tfrac{1}{2}) \, |m_1,m_2;\dot{m}_1,\dot{m}_2\rangle \ ,
\end{eqnarray}
while $J^\pm_0$ and $\hat{J}^\pm_0$ act as 
\begin{eqnarray}
J^+_0 \, |m_1,m_2;\dot{m}_1,\dot{m}_2\rangle & =  & 2\, m_1 \, |m_1-\tfrac{1}{2},m_2+\tfrac{1}{2};\dot{m}_1,\dot{m}_2\rangle \\ 
J^-_0 \, |m_1,m_2;\dot{m}_1,\dot{m}_2\rangle & =  &  2\, m_2 \, |m_1+\tfrac{1}{2},m_2-\tfrac{1}{2};\dot{m}_1,\dot{m}_2\rangle \\ 
\hat{J}^+_0 \, |m_1,m_2;\dot{m}_1,\dot{m}_2\rangle & =  & - 2\, \dot{m}_2 \, |m_1,m_2;\dot{m}_1+\tfrac{1}{2},\dot{m}_2-\tfrac{1}{2}\rangle \\ 
\hat{J}^-_0 \, |m_1,m_2;\dot{m}_1,\dot{m}_2\rangle & =  &  - 2\, \dot{m}_1 \, |m_1,m_2;\dot{m}_1-\tfrac{1}{2},\dot{m}_2+\tfrac{1}{2}\rangle \ . 
\end{eqnarray}
The Casimir 
\be
C^{\mathfrak{su}(2)} =  J^3_0 J^3_0 + \frac{1}{2} \bigl( J^+_0 J^-_0 + J^-_0 J^+_0 \bigr) =  j (j +1) 
\ee
as well as the analogously defined Casimir $\hat{C}^{\mathfrak{su}(2)}=\hat{\jmath} (\hat{\jmath}+1)$ for the other $\mathfrak{sl}(2,\mathds{R})$ algebra then takes the value 
\begin{eqnarray}
C^{\mathfrak{su}(2)} & = & j (j+1) = (m_1+m_2) (m_1+m_2+1) \ , \label{Ceig}\\
\hat{C}^{\mathfrak{su}(2)} & = & \hat{\jmath} (\hat{\jmath}+1) = (\dot{m}_1+\dot{m}_2) (\dot{m}_1+\dot{m}_2+1) \ . \label{Cheig}
\end{eqnarray}
For the $\mathfrak{su}(4)$ generators it follows from (\ref{fermizero}) that 
\be
(\psi^\dagger_b\, \psi^a)_0 \,  |m_1,m_2;\dot{m}_1,\dot{m}_2\rangle = 0 \ , \qquad 
V_0 \,  |m_1,m_2;\dot{m}_1,\dot{m}_2\rangle = - 2   \,  |m_1,m_2;\dot{m}_1,\dot{m}_2\rangle  \ . 
\ee
Finally, the eigenvalues of ${\cal B}_0$, ${\cal C}_0$, and ${\cal D}_0$ on these states are 
\begin{eqnarray}
{\cal B}_0 \, |m_1,m_2;\dot{m}_1,\dot{m}_2\rangle & =  & \bigl( m_1 + m_2 - \dot{m}_1 - \dot{m}_2 \bigr) \, |m_1,m_2;\dot{m}_1,\dot{m}_2\rangle \\ 
{\cal C}_0 \, |m_1,m_2;\dot{m}_1,\dot{m}_2\rangle & =  & \bigl( m_1 + m_2 - \dot{m}_1 - \dot{m}_2 -1 \bigr) \, |m_1,m_2;\dot{m}_1,\dot{m}_2\rangle   \label{C0R0} \\
{\cal D}_0 \, |m_1,m_2;\dot{m}_1,\dot{m}_2\rangle & =  &  \bigl( m_1 + m_2 + \dot{m}_1 + \dot{m}_2 + 1 \bigr) \, |m_1,m_2;\dot{m}_1,\dot{m}_2\rangle  \label{D0R0} \ . 
\end{eqnarray}
Again, for ${\cal B}_0$ this is a matter of convention, while the values for ${\cal C}_0$ and ${\cal D}_0$ are fixed by the (anti)-commutation relations of eqs.~(\ref{KP}) -- (\ref{dotSdotQ}). 

Using the explicit form of the Casimir of  $\mathfrak{u}(2,2|4)$, see eq.~(D.12) of \cite{Beisert:2004ry}, which in our conventions takes the form 
\begin{eqnarray}
{\cal J}^2 & = & \tfrac{1}{2} {\cal D}_0^2 + \tfrac{1}{2} {\cal L}^\gamma{}_\delta  \, {\cal L}^\delta{}_\gamma 
+ \tfrac{1}{2} \dot{\cal L}^{\dot{\gamma}}{}_{\dot{\delta}}  \, \dot{\cal L}^{\dot{\delta}}{}_{\dot{\gamma}} - \tfrac{1}{2} {\cal R}^c{}_d {\cal R}^d{}_c \\
& & - \tfrac{1}{2} [{\cal Q}^c{}_\gamma,{\cal S}^\gamma{}_c]  + \tfrac{1}{2} [\dot{\cal Q}^{\dot{\gamma}}{}_{c},\dot{\cal S}^{c}{}_{\dot{\gamma}}] + \tfrac{1}{2} \{ {\cal P}^{\dot{\gamma}}{}_{\delta},{\cal K}^{\delta}{}_{\dot{\gamma}} \}  +  {\cal B}_0 {\cal C}_0 \ , 
\end{eqnarray} 
 it is not difficult to work out its value on the state $ |m_1,m_2;\dot{m}_1 , \dot{m}_2 \rangle$, and one finds 
\be\label{Cas}
{\cal J}^2 \, |m_1,m_2;\dot{m}_1 , \dot{m}_2 \rangle = \tfrac{5}{2} \, {\cal C}_0^2 \, |m_1,m_2;\dot{m}_1 , \dot{m}_2 \rangle \ .
\ee
Thus the Casimir vanishes, ${\cal J}^2=0$, on representations of $\mathfrak{psu}(2,2|4)$ for which ${\cal C}_0=0$. 
This is also what one should have expected since we have $8$ symplectic bosons and 8 fermions, and thus in the Ramond sector the ground state energy should be zero.

\section{The single trace spectrum of 4d SYM}\label{app:B}

In this Appendix we spell out the low-lying states of the single trace spectrum of 4d SYM, following \cite{Beisert:2004di}. We then explicitly describe the correspondence between these and the physical states in the worldsheet description. 

\subsection{The complete low lying states}

As explained in \cite{Beisert:2004di}, the single trace spectrum of 4d SYM is obtained from the tensor powers of the singleton representation upon picking out the cyclically invariant combinations --- this is what the selection rules of \cite[eq.~(3.6)]{Beisert:2004di} amount to, i.e.\ the multiplicities are just the branching rules for the multiplicity of the trivial representation with respect to the cyclic subgroup ${\mathbb Z}_N \subseteq S_N$. Using the explicit form of the singleton representation, see eq.~(\ref{singleton}), we can work out the $\mathfrak{su}(2) \oplus \mathfrak{su}(2) \oplus \mathfrak{su}(4)$ representations that appear.\footnote{In doing this calculation we need to be aware of the fact that spacetime fermions contribute with a minus sign to the characters of the form ${\cal Z}_{0}(t^n)$ where ${\cal Z}_0$ is the character of ${\cal R}_0$, and $n$ is even; this is as, e.g., in the discussion of Section~3 of \cite{Gaberdiel:2015wpo}.} In the following we shall describe our results for small values of $w$ case by case. 

\subsubsection{$w=2$}

At $w=2$ there is only one Young diagram ${\tiny \yng(2)}$ that contributes,  and it leads to the states 
\begin{eqnarray}
{\tiny \yng(2)} & : &  t^2 \Bigl( (0,0;[0,2,0])^{(0)} \oplus (0,0;[0,0,0])^{(-2)} \Bigr) \nonumber \\
& & + t^{\frac{5}{2}} \Bigl( (\tfrac{1}{2},0;[0,1,1])^{(0)} \oplus  (\tfrac{1}{2},0;[1,0,0])^{(-2)} \oplus (0,\tfrac{1}{2};[1,1,0])^{(0)} \oplus  (0,\tfrac{1}{2};[0,0,1])^{(-2)} \Bigr) \nonumber  \\
& & + t^{3} \Bigl( (\tfrac{1}{2},\tfrac{1}{2};[0,2,0])^{(0)} \oplus 2 \cdot (1,0;[0,1,0])^{(0/-2)} \oplus 2 \cdot (0,1;[0,1,0])^{(0/-2)}  \nonumber  \\
& & \qquad \oplus\, 
(0,0;[0,0,2])^{(0)} \oplus (0,0;[2,0,0])^{(0)} \nonumber \\
& & \qquad \oplus \ 2 \cdot (\tfrac{1}{2},\tfrac{1}{2};[1,0,1])^{(0/-2)} \oplus   2 \cdot (\tfrac{1}{2},\tfrac{1}{2};[0,0,0]) \Bigr)  +  {\cal O}(t^{7/2}) \ ,
\end{eqnarray}
where we have organised the states by their ${\cal D}_0$ eigenvalue, which appears as the exponent of $t$, and the upper index $(n)$ denotes the $L_0=n$ eigenvalue where the corresponding representation appears in the worldsheet description.\footnote{We have written this out for those states that we have checked explicitly; for $w=2$, we also have a general argument that accounts for all states, see \cite{Gaberdiel:2021iil} and eq.~(\ref{hspin}).} We can also organise the spectrum in terms of $\mathfrak{psu}(2,2|4)$ representations, in which case the answer has the simple form, see \cite{Beisert:2004di}
\be
{\tiny \yng(2)} : \qquad \bigl[ 0,0;[0,2,0] \bigr]_{2}^{(0)} \oplus \bigoplus_{p=1}^{\infty} \,  \bigl[ p-1,p-1;[0,0,0] \bigr]_{2p}^{(-2p)} \ . 
\ee
The first term is the BPS multiplet with the stress tensor. The $p=1$ term is the Konishi multiplet with 
${\rm Tr}[\phi^i\phi^i]$ as its highest weight state. The $p>1$ terms are higher spin conserved current multiplets (of twist two), and are therefore short representations of $\mathfrak{psu}(2,2|4)$ in the free theory. 

\subsubsection{$w=3$}

At $w=3$ the low lying terms are 
\begin{eqnarray}
{\tiny \yng(3)} & : & t^3 \Bigl( (0,0;[0,3,0])^{(0)} \oplus (0,0;[0,1,0])^{(-3)} \Bigr) \nonumber  \\
& & + t^{\frac{7}{2}} \Bigl( (\tfrac{1}{2},0;[0,2,1])^{(0)} \oplus (0,\tfrac{1}{2};[1,2,0])^{(0)} \oplus (\tfrac{1}{2},0;[1,1,0]) \oplus (0,\tfrac{1}{2};[0,1,1]) \nonumber \\
& & \qquad  \oplus\ (\tfrac{1}{2},0;[0,0,1]) \oplus (0,\tfrac{1}{2};[1,0,0]) \Bigr)  \nonumber \\
& & + t^4 \Bigl( (\tfrac{1}{2},\tfrac{1}{2};[0,3,0])^{(0)} \oplus 2\cdot (1,0;[0,2,0])^{(0/-3)} \oplus 2\cdot (0,1;[0,2,0])^{(0/-3)}  \nonumber \\
& & \qquad \oplus \  (0,0;[2,1,0])^{(0)} \oplus (0,0;[0,1,2])^{(0)} \oplus 2 \cdot (\tfrac{1}{2},\tfrac{1}{2};[1,1,1]) \nonumber \\
& & \qquad \oplus\  (1,0;[1,0,1]) \oplus (0,1;[1,0,1]) \nonumber \\
& & \qquad \oplus\ (\tfrac{1}{2},\tfrac{1}{2};[2,0,0]) \oplus (\tfrac{1}{2},\tfrac{1}{2};[0,0,2]) \oplus 4\cdot (\tfrac{1}{2},\tfrac{1}{2};[0,1,0]) \oplus 2\cdot (0,0;[1,0,1]) \nonumber \\
& & \qquad \oplus\ 2\cdot (1,0;[0,0,0]) \oplus 2\cdot (0,1;[0,0,0]) \Bigr) + {\cal O}(t^{9/2}) \ ,
\end{eqnarray}
and
\begin{eqnarray}
 {\tiny \yng(1,1,1)} & : & t^3 \Bigl( (0,0;[2,0,0])^{(-3)} \oplus (0,0;[0,0,2])^{(-3)} \Bigr) \nonumber  \\
 & & + t^{\frac{7}{2}} \Bigl( (\tfrac{1}{2},0;[1,1,0]) \oplus (0,\tfrac{1}{2};[0,1,1]) \oplus 
 (\tfrac{1}{2},0;[1,0,2]) \oplus (0,\tfrac{1}{2};[2,0,1]) \nonumber \\
 & & \qquad + (\tfrac{1}{2},0;[0,0,1]) \oplus (0,\tfrac{1}{2};[1,0,0]) \Bigr) \nonumber \\ 
 & & + t^{4} \Bigl( 2 \cdot (0,0;[0,2,0])^{(-3)}  \oplus  (1,0;[0,1,2])^{(-3)} \oplus (0,1;[2,1,0])^{(-3)}    \nonumber \\
& & \qquad \oplus\ 2 \cdot (\tfrac{1}{2},\tfrac{1}{2};[1,1,1]) \oplus 2\cdot (1,0;[1,0,1]) \oplus 2\cdot (0,1;[1,0,1]) 
\nonumber \\
& & \qquad \oplus\  2\cdot (\tfrac{1}{2},\tfrac{1}{2};[2,0,0]) \oplus 2\cdot (\tfrac{1}{2},\tfrac{1}{2};[0,0,2]) \oplus
3\cdot (\tfrac{1}{2},\tfrac{1}{2};[0,1,0]) \nonumber \\
& & \qquad \oplus\ 2\cdot (0,0;[1,0,1]) \oplus 2\cdot (0,0;[0,0,0]) \Bigr) +  {\cal O}(t^{9/2}) \ ,
\end{eqnarray}
where the exponent of $t$ is again the ${\cal D}_0$ eigenvalue, and the upper index $(n)$ denotes the $L_0=n$ eigenvalue at which the corresponding state appears in the worldsheet analysis. In terms of $\mathfrak{psu}(2,2|4)$ representations, the low-lying states are 
\be\label{B.5}
{\tiny \yng(3)}: \qquad \underbrace{\bigl[ 0,0;[0,3,0] \bigr]_{3}}_{{\cal V}_{0,0}}{}^{\!\!\!\! (0)} \oplus  
\underbrace{\bigl[ 0,0;[0,1,0] \bigr]_{3}}_{{\cal V}_{0,2}}{}^{\!\!\!\!  (-3)} \oplus \cdots
\ee
and
\be\label{B.6}
{\tiny \yng(1,1,1)}: \qquad \underbrace{\bigl[ 0,0;[2,0,0] \bigr]_{3}}_{{\cal V}_{-1,0}}{}^{\!\!\!\! (-3)} \oplus 
\underbrace{\bigl[ 0,0;[0,0,2] \bigr]_{3}}_{{\cal V}_{1,0}}{}^{\!\!\!\! (-3)}  \oplus \cdots \ , 
\ee
where the ${\cal V}_{n,k}$ representations are defined as in \cite{Beisert:2004di}, see eq.~(3.8), Table~1, and eq.~(A.8).

\subsection{Fermionic Subsectors}\label{app:Cferm}

We have also worked out the states with ${\cal D}_0 = w$; they sit necessarily in the singlet representation $(0,0)$ with respect to $\mathfrak{su}(2) \oplus \mathfrak{su}(2)$, and their $\mathfrak{su}(4)$ representation content is as follows:
\begin{equation}
\begin{array}{rrl}
w=2: \qquad & {\tiny \yng(2)}: \quad  & [0,2,0]^{(0)} \oplus [0,0,0]^{(-2)} \\[4pt]
w=3: \qquad & {\tiny \yng(3)}: \quad  &  [0,3,0]^{(0)} \oplus [0,1,0]^{(-3)} \\
& {\tiny \yng(1,1,1)}: \qquad & [2,0,0]^{(-3)} \oplus [0,0,2]^{(-3)} \\[4pt]
w=4: \qquad & {\tiny \yng(4)}: \quad & [0,4,0]^{(0)} \oplus [0,2,0]^{(-4)} \oplus [0,0,0]^{(-8)} \\[2pt]
& {\tiny \yng(2,2)}: \qquad & [2,0,2]^{(-4)} \oplus [0,2,0]^{(-4)} \oplus [0,0,0]^{(-8)}  \\[4pt]
& {\tiny \yng(2,1,1)}: \qquad & [2,1,0]^{(-4)} \oplus [0,1,2]^{(-4)} \oplus [1,0,1]^{(-4)} \\[4pt]
w=5: \qquad & {\tiny \yng(5)}: \quad & [0,5,0]^{(0)} \oplus [0,3,0]^{(-5)} \oplus [0,1,0] \\
& {\tiny \yng(3,2)}: \qquad & [0,3,0]^{(-5)} \oplus [2,1,2]^{(-5)} \oplus [1,1,1]^{(-5/10)}  \oplus [0,1,0]\\[2pt]
& 2\cdot {\tiny \yng(3,1,1)}: \quad & [2,2,0]^{(-5)} \oplus [0,2,2]^{(-5)} \oplus [1,1,1]^{(-5/10)} \oplus [2,0,0] \oplus [0,0,2]
\\[6pt]
& {\tiny \yng(2,2,1)}: \quad & [1,1,1]^{(-5/10)} \oplus [3,0,1]^{(-5)} \oplus [1,0,3]^{(-5)} \oplus [0,1,0]\\[6pt]
& {\tiny \yng(1,1,1,1,1)}: \quad & [0,1,0] \ ,
\end{array}
\end{equation}
where the multiplicity $2\cdot $ in ${\tiny \yng(3,1,1)}$ is to remind us that these states appear with multiplicity $2$ (since the coresponding representation of $S_5$ contains the trivial representation of $\mathbb{Z}_5 \subset S_5$ twice). Furthermore, the upper index $(n)$ denotes the $L_0=n$ eigenvalue at which the corresponding state appears in the worldsheet analysis. For $w=5$, of the $4$ states in the representation $[1,1,1]$, $3$ of them appear at $L_0=-5$, while one appears at $L_0=-10$. 

\subsection{Bosonic Subsectors}\label{app:Cbos}

We have also worked out the states that transform in the $[0,w,0]$ representation of $\mathfrak{su}(4)$; they arise from purely bosonic descendants. For them we only describe the $\mathfrak{su}(2) \oplus \mathfrak{su}(2)$ quantum numbers, as well as the eigenvalue of ${\cal D}_0$ (as the exponent of~$t$), and the low-lying states organise themselves as follows.
\subsubsection{$w=2$}
\be\label{w2bos}
\begin{array}{lcl}
{\tiny \yng(2)} : & &  t^2 (0,0)^{(0)} + t^3 (\tfrac{1}{2},\tfrac{1}{2})^{(0)}  + t^4 \Bigl( 2\cdot (1,1)^{(0/-2)} \oplus (0,0)^{(0)} \Bigr) \\
& & + t^5 \Bigl( 2 \cdot (\tfrac{3}{2},\tfrac{3}{2})^{(0/-2)} \oplus (\tfrac{1}{2},\tfrac{3}{2})^{(-2)}  \oplus (\tfrac{3}{2},\tfrac{1}{2})^{(-2)} \oplus (\tfrac{1}{2},\tfrac{1}{2})^{(0)} \Bigr) + {\cal O}(t^6) \ .
\end{array}
\ee

\subsubsection{$w=3$}
\be
\begin{array}{lcl}
{\tiny \yng(3)}: & & t^3 (0,0)^{(0)}  + t^4 (\tfrac{1}{2},\tfrac{1}{2})^{(0)}  + t^5 \Bigl( 2\cdot (1,1)^{(0/-3)} \oplus (0,0)^{(0)} \Bigr) \\ 
& & + t^6 \Bigl( 3 \cdot (\tfrac{3}{2},\tfrac{3}{2}) \oplus (\tfrac{1}{2},\tfrac{3}{2})^{(-3)} \oplus (\tfrac{3}{2},\tfrac{1}{2})^{(-3)} \oplus 2\cdot (\tfrac{1}{2},\tfrac{1}{2})^{(0/-3)} \Bigr) + {\cal O}(t^7) \\[8pt]
{\tiny \yng(1,1,1)}: & & t^5 \Bigl( (1,0)^{(-3)} \oplus (0,1)^{(-3)} \Bigr) 
+ t^6 \Bigl((\tfrac{3}{2},\tfrac{3}{2}) \oplus (\tfrac{1}{2},\tfrac{3}{2})^{(-3)} \oplus (\tfrac{3}{2},\tfrac{1}{2})^{(-3)} \oplus 2\cdot (\tfrac{1}{2},\tfrac{1}{2})^{(-3)} \Bigr) \\[2pt]
& & + {\cal O}(t^7) \ ,
\end{array}
\ee
where of the $4$ states with $t^6 (\tfrac{3}{2},\tfrac{3}{2})$, 1 arises at $L_0=0$, 2 arise at $L_0=-3$, and 1 at $L_0=-6$.

\subsubsection{$w=4$}
\be
\begin{array}{lcl}
{\tiny \yng(4)}: & &  t^4(0,0)^{(0)}  + t^5 (\tfrac{1}{2},\tfrac{1}{2})^{(0)}  + t^6 \Bigl( 2\cdot (1,1)^{(0/-4)} \oplus (0,0)^{(0)} \Bigr) \\
& & + t^7 \Bigl( 3 \cdot (\tfrac{3}{2},\tfrac{3}{2}) \oplus (\tfrac{1}{2},\tfrac{3}{2}) \oplus (\tfrac{3}{2},\tfrac{1}{2}) \oplus 2\cdot (\tfrac{1}{2},\tfrac{1}{2})^{(0/-4)} \Bigr) +  {\cal O}(t^8) \\[8pt]
{\tiny \yng(2,2)}:  & & t^6 \Bigl( (1,1)^{(-4)} \oplus (0,0)^{(-4)} \Bigr) 
+ t^7 \Bigl( (\tfrac{3}{2},\tfrac{3}{2}) \oplus  2\cdot (\tfrac{1}{2},\tfrac{3}{2}) \oplus  2\cdot (\tfrac{3}{2},\tfrac{1}{2}) \oplus 2\cdot (\tfrac{1}{2},\tfrac{1}{2})^{(-4)} \Bigr)\\[4pt]
& & + {\cal O}(t^8) \\[8pt]
{\tiny \yng(2,1,1)}: & & t^6 \Bigl( (1,0)^{(-4)} \oplus (0,1)^{(-4)} \Bigr) 
+ t^7 \Bigl( (\tfrac{3}{2},\tfrac{3}{2}) \oplus  2\cdot (\tfrac{1}{2},\tfrac{3}{2}) \oplus  2\cdot (\tfrac{3}{2},\tfrac{1}{2}) \oplus 3\cdot (\tfrac{1}{2},\tfrac{1}{2})^{(-4)} \Bigr) \\[4pt]
& &  + {\cal O}(t^8) \ ,
\end{array}
\ee
where of the $5$ states with $t^7 (\tfrac{3}{2},\tfrac{3}{2})$, 1 arises at $L_0=0$, 3 arise at $L_0=-4$, and 1 at $L_0=-8$, while of the $5$ states in $t^7 (\tfrac{3}{2},\tfrac{1}{2})$ or $t^7 (\tfrac{1}{2},\tfrac{3}{2})$, $4$ arise at $L_0=-4$, while $1$ arises at $L_0=-8$.

\section{The analysis for ${\rm AdS}_3$}\label{app:C}

In this Appendix we perform the physical wedge state analysis for the case of ${\rm AdS}_3$. We begin by reviewing our conventions following \cite{Dei:2020zui}. 

\subsection{The ${\rm AdS}_3$ free fields}\label{app:AdS3}

In the case of ${\rm AdS}_3$ the free field worldsheet fields are a pair of symplectic bosons $(\xi^\pm,\eta^\pm)$ and complex fermions $(\psi^\pm,\chi^\pm)$ with commutation relations 
\be
{} \{ \psi^\alpha_r,\chi^\beta_s\} = \epsilon^{\alpha\beta} \, \delta_{r,-s} \ , \qquad 
[\xi^\alpha_r,\eta^\beta_s] = \epsilon^{\alpha\beta}\, \delta_{r,-s} \ . 
\ee
These modes generate the algebra $\mathfrak{u}(1,1|2)_1$, 
\begin{subequations}\label{u112}
\begin{align}
J^3_m&=-\tfrac{1}{2} (\eta^+\xi^-)_m-\tfrac{1}{2} (\eta^-\xi^+)_m\ , & K^3_m&=-\tfrac{1}{2} (\chi^+\psi^-)_m-\tfrac{1}{2} (\chi^-\psi^+)_m\ , \\
J^\pm_m&=(\eta^\pm\xi^\pm)_m\ , & K^\pm_m&=\pm(\chi^\pm\psi^\pm)_m\ , \\
S_m^{\alpha\beta+}&=(\chi^\beta \xi^\alpha)_m\ , & S_m^{\alpha\beta-}&=-(\eta^\alpha\psi^\beta)_m \ ,  \\
U_m&=-\tfrac{1}{2} (\eta^+\xi^-)_m+\tfrac{1}{2} (\eta^-\xi^+)_m\ , & V_m&=-\tfrac{1}{2} (\chi^+\psi^-)_m+\tfrac{1}{2} (\chi^-\psi^+)_m\ .
\end{align}\label{eq:u112-algebra}
\end{subequations}
In order to obtain $\mathfrak{psu}(1,1|2)$ from $\mathfrak{u}(1,1|2)_1$, we need to quotient out by the modes 
$Z_m = U_m + V_m$. Spectral flow is defined by\footnote{We have translated the `active' conventions of \cite{Dei:2020zui} to the `passive' conventions being used in this paper, see also Footnote~\ref{foot1}.} 
\be\label{AdS3sf}
\tilde{\xi}^\pm_r =  \xi^\pm_{r\pm \frac{w}{2}} \ ,  \qquad 
\tilde{\eta}^\pm_r =  \eta^\pm_{r\pm \frac{w}{2}} \ ,  \qquad 
\tilde{\psi}^\pm_r =  \psi^\pm_{r\mp \frac{w}{2}} \ , \qquad 
\tilde{\chi}^\pm_r =  \chi^\pm_{r\mp \frac{w}{2}} \ ,
\ee
and this implies that for the bosonic generators of $\mathfrak{u}(1,1|2)$ we find (among others)
\be\label{AdS3sfC}
\begin{array}{rclrcl} 
\tilde{J}^3_n & = & J^3_n - \tfrac{w}{2}\, \delta_{n,0}   \qquad  & \tilde{K}^3_n & = & K^3_n - \tfrac{w}{2}\, \delta_{n,0} \\
\tilde{U}_m & = & U_n  \qquad & \tilde{V}_n & = & V_n \\ 
\tilde{L}_n & = & L_n - w  (K^3_n - J^3_n)  \ , \qquad & \tilde{Z}_n & = & Z_n \ . 
\end{array}
\ee

The basic representation from which we shall start is the NS vacuum representation which is generated from a ground state satisfying 
\be
\eta^\pm_r \, |0\rangle = \xi^\pm_r \, |0\rangle = \chi^\pm_r \, |0\rangle = \psi^\pm_r \, |0\rangle = 0 \qquad r \geq \tfrac{1}{2} \ . 
\ee
Note that this state has the property that also 
\be
U_0 \, |0\rangle = V_0 \, |0\rangle = Z_0\,  |0\rangle = 0 \ . 
\ee
As for the ${\rm AdS}_5$ case discussed in the main part of the paper, the $w=1$ spectrally flowed NS-representation is the R-representation, and in the conventions of \cite{Dei:2020zui} we have 
\be\label{AdS3w1}
|0\rangle_1 = |\tfrac{1}{2},0\rangle \ . 
\ee
Here we have used that, see in particular eq.~(2.13) and (2.14) of \cite{Dei:2020zui}
\begin{eqnarray} \label{eigenvs}
J^3_0 \, |m_1,m_2\rangle & =& (m_1+m_2) \, |m_1,m_2\rangle \label{J30} \\
U_0 \, |m_1,m_2\rangle & =& (m_1-m_2 - \tfrac{1}{2}) \, |m_1,m_2\rangle  \ , \label{U0}
\end{eqnarray}
as well as \cite[eq.~(2.17)]{Dei:2020zui}
\be
\chi^+_0 \, |m_1,m_2 \rangle = \psi^+_0 \, |m_1,m_2 \rangle = 0 \ , \qquad K^3_0 \, |m_1,m_2 \rangle = + \tfrac{1}{2}\, |m_1,m_2 \rangle \ . 
\ee

\subsection{The physical wedge states for $w=3$}\label{app:AdS3w3}

By the same argument as for ${\rm AdS}_5$ the physical wedge states are counted by 
\begin{eqnarray}
Z^{(3)}_{{\tiny \yng(3)}}(t,y) & = &  \frac{1}{6} \Bigl(  \bigl( Z^{(1)}(t,y) \bigr)^3 + 3 \, \tilde{Z}^{(1)}(t^2,y^2)  Z^{(1)}(t,y) + 2 Z^{(1)}(t^3,y^3) \Bigr) \\ 
Z^{(3)}_{{\tiny \yng(1,1,1)}}(t,y) & = &  \frac{1}{6} \Bigl(  \bigl( Z^{(1)}(t,y) \bigr)^3 - 3 \, \tilde{Z}^{(1)}(t^2,y^2)  Z^{(1)}(t,y) + 2 Z^{(1)}(t^3,y^3) \Bigr) \ , 
\end{eqnarray}
where $\tilde{Z}^{(1)}$ was defined in (\ref{Z1tilde}), and we have also verified this to low levels. If we define the ${\cal N}=4$ BPS character (this generalises (\ref{N4vac}), to which it reduces for $h=0$) via 
\be
\chi_{h}^{{\rm (BPS)}} (t,y)  = \frac{t^{h+1}}{1-t} \Bigl( \chi_{h+1}(y)  + 2 \, t^{1/2} \, \chi_{h+\frac{1}{2}}(y) + t \,\chi_{h}(y) \Bigr) \ , 
\ee
then we can write $Z^{(3)}_{{\tiny \yng(3)}}(t,y)$ and $Z^{(3)}_{{\tiny \yng(1,1,1)}}(t,y)$ as 
\begin{eqnarray}\label{repdecomp}
Z^{(3)}_{{\tiny \yng(3)}}(t,y) & = &   \chi_{\frac{1}{2}}^{{\rm (BPS)}} (t,y) + 
\sum_{n=2}^{\infty} c_{n} \, \chi_{\frac{1}{2}}(y)  \, \chi_{\frac{1}{2}+n}^{\,{\cal N}=4}(t,y) + 
2 \sum_{n=0}^{\infty} c_n \, \chi_{5+n}^{\,{\cal N}=4}(t,y) \nonumber \\
Z^{(3)}_{{\tiny \yng(1,1,1)}}(t,y) & = &  2 \sum_{n=0}^{\infty} c_n\, \chi_{2+n}^{\, {\cal N}=4}(t,y)
+ \sum_{n=0}^{\infty}  c_n\, \chi_{\frac{1}{2}}(y) \, \chi_{\frac{7}{2}+n}^{\, {\cal N}=4}(t,y)  \ ,
\end{eqnarray}
where $c_n$ are the expansion coefficients in the power series, see \cite[eq.~(4.23)]{Beisert:2004di}
\be
\sum_{n=0}^{\infty} c_n x^n = \frac{1}{(1-x^2)(1-x^3)} \ , 
\ee
and are explicitly given as 
\be\label{cdef}
c_n = 1 + \lfloor \tfrac{n}{6} \rfloor - \delta_{n,1}^{(6)}  \ , \qquad \quad \delta_{n,1}^{(6)} = \left\{ \begin{array}{cl} 1 \quad & \hbox{if $n=1$ mod $6$} \\ 0 \quad & \hbox{otherwise.} \end{array} \right. 
\ee
As before, the $L_0=0$ contribution just accounts for the BPS multiplet with character $\chi_{\frac{1}{2}}^{{\rm (BPS)}} (t,y)$. In order to understand which states contribute at $L_0=-3p$ let us introduce as in eq.~(\ref{Sdef}) the generators 
\be
S^{\eta\xi}_2 = \eta^+_{1}\, \xi^+_{1}  \ , \qquad S^{\eta\xi}_1 = \eta^+_{1} \xi^+_{0} + \eta^+_{0} \xi^+_{1}  \ , 
\ee
and similarly for $S^{\chi\psi}_n$, $S^{\chi\xi}_n$ and $S^{\eta\psi}_n$. These generators satisfy a number of non-trivial relations, e.g.\ 
\begin{eqnarray}
S^{\eta \xi}_2 \, S^{\chi\psi}_2 & = & S^{\chi \xi}_2 \, S^{\eta\psi}_2 \\
S^{\eta\xi}_2 S^{\chi\psi}_1 + S^{\eta\xi}_1 S^{\chi\psi}_2 & = & S^{\chi\xi}_2 S^{\eta\psi}_1 + S^{\chi\xi}_1 S^{\eta\psi}_2 \ . 
\end{eqnarray}
Then the physical highest weight states at $L_0 =  - 6m$ (with $m\geq 1$) are in one-to-one correspondence to
\begin{eqnarray}
\bigl(S^{\eta\xi}_2 \bigr)^{3m-1} \, S^{\chi\psi}_2 |0\rangle_3 \quad & \longleftrightarrow & \quad 
\chi_{\frac{1}{2}}(y) \, \chi_{3m+ \frac{1}{2}}^{\, {\cal N}=4}(t,y) \ ,  \label{6mstates} \\ 
\bigl(S^{\eta\xi}_2 \bigr)^{3m-r} \, \bigl(S^{\eta\xi}_1 \bigr)^{2r-1} \, S^{\chi\psi}_1 |0\rangle_3 \quad & \longleftrightarrow & \quad 
\chi_{\frac{1}{2}}(y) \, \chi_{3m + \frac{1}{2}+r}^{\, {\cal N}=4}(t,y) \ , \ \ (r=1,\ldots,3m) \ , \nonumber \\ 
\bigl(S^{\eta\xi}_2 \bigr)^{3m-2-r} \, \bigl(S^{\eta\xi}_1 \bigr)^{2r+1} \, S^{\chi\xi}_1 S^{\chi\psi}_2 |0\rangle_3 \quad & \longleftrightarrow & \quad 
 \chi_{3m +r+1}^{\, {\cal N}=4}(t,y) \ , \ \ (r=0,\ldots,3m-2) \ , \nonumber \\
\bigl(S^{\eta\xi}_2 \bigr)^{3m-2-r} \, \bigl(S^{\eta\xi}_1 \bigr)^{2r+1} \, S^{\eta\psi}_1 S^{\chi\psi}_2 |0\rangle_3 \quad & \longleftrightarrow & \quad 
 \chi_{3m + r+1}^{\, {\cal N}=4}(t,y) \ , \ \ (r=0,\ldots,3m-2) \ , \nonumber 
\end{eqnarray}
while those at $L_0 =  - 6m-3$ are
\begin{eqnarray}
\bigl(S^{\eta\xi}_2 \bigr)^{3m+1-r} \, \bigl(S^{\eta\xi}_1 \bigr)^{2r} \, S^{\chi\psi}_1 |0\rangle_3 \quad & \longleftrightarrow & \quad 
\chi_{\frac{1}{2}}(y) \, \chi_{3m + \frac{5}{2}+r}^{\, {\cal N}=4}(t,y) \ , \ \ (r=0,\ldots,3m+1) \ , \nonumber \\ 
\bigl(S^{\eta\xi}_2 \bigr)^{3m-r} \, \bigl(S^{\eta\xi}_1 \bigr)^{2r} \, S^{\chi\xi}_1 S^{\chi\psi}_2 |0\rangle_3 \quad & \longleftrightarrow & \quad 
 \chi_{3m +r+2}^{\,{\cal N}=4}(t,y) \ , \ \ (r=0,\ldots,3m) \ , \nonumber \\
\bigl(S^{\eta\xi}_2 \bigr)^{3m-r} \, \bigl(S^{\eta\xi}_1 \bigr)^{2r} \, S^{\eta\psi}_1 S^{\chi\psi}_2 |0\rangle_3 \quad & \longleftrightarrow & \quad 
 \chi_{3m + r+2}^{\, {\cal N}=4}(t,y) \ , \ \ (r=0,\ldots,3m) \ . \label{6m+3states} 
\end{eqnarray}
Thus the complete contributions at $L_0=-6 m$ and $L_0=-6 m -3$ are 
\begin{eqnarray}
Z^{(3)}_{L_0=-6 m } (t,y) & = &  \sum_{r=0}^{3m} \chi_{\frac{1}{2}}(y) \, \chi_{3m + \frac{1}{2} + r }^{\, {\cal N}=4}(t,y)   + 2 \, \sum_{r=0}^{3m-2} \chi_{3m + 1 + r }^{\, {\cal N}=4}(t,y) \ , \label{L6m}\\ 
Z^{(3)}_{L_0=-6 m-3 } (t,y) & = &  \sum_{r=0}^{3m+1} \chi_{\frac{1}{2}}(y) \, \chi_{3m + \frac{5}{2} + r }^{\, {\cal N}=4}(t,y)   + 2 \, \sum_{r=0}^{3 m } \chi_{3m + 2 + r }^{\, {\cal N}=4}(t,y) \ ,  \label{L6mp3}
\end{eqnarray}
and we verify that indeed 
\be
Z^{(3)}_{{\tiny \yng(3)}}(t,y) + Z^{(3)}_{{\tiny \yng(1,1,1)}}(t,y) = \chi_{\frac{1}{2}}^{{\rm (BPS)}} (t,y) +  \sum_{m=0}^{\infty} \Bigl[
Z^{(3)}_{L_0=-6 m-6 } (t,y) + Z^{(3)}_{L_0=-6 m-3 } (t,y) \Bigr] \ . 
\ee
Comparing this with with eq.~\eqref{repdecomp} we see that the worldsheet gives a refined and more natural organisation of the physical states. 
 
\subsection{The even spin ${\cal W}_\infty$ algebra}\label{app:Winf}

In this appendix we describe the structure of the even spin ${\cal W}_\infty$ algebra that is captured by the physical wedge states. We begin with explaining the structure of the ${\cal W}_\infty$ algebra associated to the symmetric orbifold of $\mathbb{T}^4$. In a complex basis, the $\mathbb{T}^4$ theory has two pairs of complex boson and fermion fields,\footnote{In the following we shall only discuss the left-moving fields. To avoid cluttering our notation, we shall also drop the index labelling the $N$ copies; a sum over all the copies is always understood.}
\be
(\partial \phi^1, \partial \phi^2) \ , \qquad (\partial \bar{\phi}^1, \partial \bar{\phi}^2) \ , \qquad \qquad 
(\psi^1, \psi^2) \ , \qquad (\bar{\psi}^1, \bar{\psi}^2) \ ,
\ee
with standard OPEs, e.g.\ 
\be
\partial \phi^i(z) \, \partial \bar\phi^j(w) \sim \frac{\delta^{ij}}{(z-w)^2} \ , \qquad 
 \psi^i(z) \,  \bar\psi^j(w) \sim \frac{\delta^{ij}}{z-w} \ . 
 \ee
The generators of the ${\cal N}=2$ superconformal algebra can be taken to be 
\begin{eqnarray}
J & = & \bar\psi^1 \psi^1 + \bar\psi^2 \psi^2 \\
G^+ & = & \sqrt{2}\, \bigl( \partial \phi^1 \bar{\psi}^1 +  \partial \phi^2 \bar{\psi}^2  \bigr) \\ 
G^- & = & \sqrt{2} \, \bigl( \partial \bar{\phi}^1 \psi^1 +  \partial \bar{\phi}^2 \psi^2 \bigr) \\ 
L & = & \tfrac{1}{2} \bigl( \partial \bar{\psi}^1 \psi^1 - \bar{\psi}^1 \partial \psi^1  +  \partial \bar{\psi}^2 \psi^2 -  \bar{\psi}^2 \partial \psi^2   \bigr) + \partial \bar{\phi}^1 \partial \phi^1 + \partial \bar{\phi}^2 \partial \phi^2 \ . 
\end{eqnarray}
They can be extended to an ${\cal N}=4$ algebra by adding in the currents 
\be
J^+ = \bar{\psi}^1 \bar{\psi}^2 \ , \qquad J^- = \psi^2 \psi^1 \ , 
\ee
as well as the additional supercurrents 
\be
G^{'+} = \sqrt{2}\, \bigl(  \bar{\psi}^1 \partial \bar{\phi}^2 - \bar{\psi}^2 \partial \bar{\phi}^1  \bigr) \ , \qquad 
G^{'-} = \sqrt{2}\, \bigl( \psi^1 \partial \phi^2 - \psi^2 \partial \phi^1 \bigr) \ . 
\ee
The ${\cal W}_\infty$ algebra that appears in the $\mathbb{T}^4$ theory is now generated by all the fields of the form 
\be
\partial^l S^i \, \partial^m \bar{S}^i \ , \ \ i=1,2 \ , \qquad \partial^l S^1 \partial^m {S}^2 \ , \qquad \partial^l \bar{S}^1 \partial^m \bar{S}^2 \ , 
\ee
where $l,m\in \mathbb{N}_0$, and $S^i$ stands for either $\psi^i$ or $\partial\phi^i$. This algebra is generated by the above ${\cal N}=4$ superconformal generators, together with the higher spin fields in ${\cal R}^{(s)}$ for $s=1,2,3,\ldots$. Here we use the notation of \cite[eq.~(2.3)]{Gaberdiel:2014cha}, and for example the lowest spin-$1$ field in ${\cal R}^{(1)}$ is 
\be\label{W1}
W^{(1)} = \bar\psi^1 \psi^1 - \bar\psi^2 \psi^2 \ . 
\ee

The higher spin ${\cal W}_\infty$ algebra that appears in the spectrum of our worldsheet description, see eq.~(\ref{w2complete}), is the subalgebra that is invariant under the automorphism 
\be
S^1 \mapsto S^2 \mapsto - S^1 \ , \qquad  \bar{S}^1 \mapsto \bar{S}^2 \mapsto - \bar{S}^1 \ . 
\ee
This leaves the above ${\cal N}=4$ generators invariant, but not for example the field $W^{(1)}$ in (\ref{W1}). It is not difficult to check that its spectrum is described by eq.~(\ref{w2complete}), i.e.\ only the ${\cal N}=4$ algebra together with the even spin multiplets ${\cal R}^{(s)}$ with $s$ even survive. (Indeed, for each half-integer conformal dimension $h=\frac{3}{2},2,\frac{5}{2},\ldots$, there are four independent (non-derivative) generators.)

\subsection{The minimal $\mathfrak{psu}(1,1|2)$ representations}\label{app:C.4}

As is well known the (super)-conformal symmetry algebra in 2d is much bigger than its higher dimensional analogue. Indeed, the former is infinite dimensional, whereas the higher dimensional (super)-conformal algebras are all finite-dimensional. The natural analogue of the higher dimensional conformal symmetry in 2d is the M\"obius group --- the group of global conformal transformations on the sphere --- whose Lie algebra is $\mathfrak{sl}(2,\mathds{R})$. Similarly, the natural analogue of the higher dimensional ${\cal N}=4$ superconformal algebra in 2d is the Lie algebra $\mathfrak{psu}(1,1|2)$. It is obtained from the small ${\cal N}=4$ superconformal algebra upon restricting to the (global) wedge modes, i.e.\ the generators $L_0$, $L_{\pm 1}$, $G^\pm_{\pm \frac{1}{2}}$, $G^{'\pm}_{\pm \frac{1}{2}}$, as well as $J^a_0$ with $a=3$ and $a=\pm$. 

The worldsheet description in terms of the WZW model preserves manifestly this $\mathfrak{psu}(1,1|2)$ algebra, since the zero modes of the affine $\mathfrak{psu}(1,1|2)$ generators map physical states to physical states.\footnote{This is also true in the approach of the present paper since the relevant symmetry generators are the $S^{ab}_0$ zero modes.} We can therefore organise all physical states in terms of representations of $\mathfrak{psu}(1,1|2)$. 

The simplest (non-trivial) representation of $\mathfrak{psu}(1,1|2)$ is the `minimal' representation with character (\ref{Z1AdS3}). Microscopically, it can be described by the $\mathfrak{psu}(1,1|2)$ descendants acting on $\bar\psi^1_{-1/2} |0\rangle_{\rm NS}$,  say, since this leads to the states
\be\label{fund1}
\bar\psi^1_{-r} |0\rangle_{\rm NS} \ , \quad \psi^2_{-r} |0\rangle_{\rm NS} \ , \qquad \bar\alpha^1_{-n} |0\rangle_{\rm NS} \ , \quad \alpha^2_{-n} |0\rangle_{\rm NS} \ , 
\ee
where $r=\frac{1}{2},\frac{3}{2},\ldots$ and $n=1,2,\ldots$, and $\alpha^i_n$ are the modes of $\partial\phi^i$. Incidentally, these states also form a representation with respect to the higher spin algebra, i.e.\ the wedge algebra corresponding to the above ${\cal W}_\infty$ algebra. Obviously, the same is true for the conjugate representation, i.e.\ the one generated by the action of $\mathfrak{psu}(1,1|2)$
acting on $\psi^1_{-1/2} |0\rangle_{\rm NS}$,  say; it leads to the states 
\be\label{fund2}
\psi^1_{-r} |0\rangle_{\rm NS} \ , \quad \bar\psi^2_{-r} |0\rangle_{\rm NS} \ , \qquad \alpha^1_{-n} |0\rangle_{\rm NS} \ , \quad \bar\alpha^2_{-n} |0\rangle_{\rm NS} \ .
\ee

The same representation of $\mathfrak{psu}(1,1|2)$  also appears in the $2$-cycle twisted sector. In the $2$-cycle twisted sector the fields $S^i$ and $\bar{S}^i$ have both integer and half-integer modes. The ground state $|0\rangle^{(2)}$ is annihilated by all positive modes, and satisfies in addition 
\be
\alpha^i_0 |0\rangle^{(2)} = \bar{\alpha}^i_0 |0\rangle^{(2)} = 0 \ , \qquad 
\bar\psi^1_0 \, |0\rangle^{(2)} =  \bar\psi^2_0\,  |0\rangle^{(2)} = 0 \ . 
\ee
In these conventions, $|0\rangle^{(2)}$ is the spin-up component of a $\mathfrak{su}(2)$ doublet --- the spin-down component is $J^-_0 |0\rangle^{(2)} = \psi^2_0 \psi^1_0 |0\rangle^{(2)}$. It has $h=\frac{1}{2}$ and it is invariant with respect to the $\mathbb{Z}_2$ centraliser. (There are also two $\mathfrak{su}(2)$ singlets, namely $\psi^i_0 |0\rangle^{(2)}$ with $i=1,2$, but they are odd under the $\mathbb{Z}_2$ centraliser.) It therefore has the same quantum numbers ($h=j=\frac{1}{2}$) as the minimal representation from above, and hence defines an equivalent representation. (This is familiar from the fact that it also leads to an exactly marginal operator.) In particular, its character is therefore also given by eq.~(\ref{Z1AdS3}). For example, at excitation level $\frac{1}{2}$, there are only two descendants, namely  
\begin{align} \label{halfdes}
G^-_{-\frac{1}{2}} |0\rangle^{(2)} & = \bigl( \bar{\alpha}^1_{-\frac{1}{2}} \psi^1_0 + \bar{\alpha}^2_{-\frac{1}{2}} \psi^2_0 \bigr) |0\rangle^{(2)} \\
G^{'-}_{-\frac{1}{2}} |0\rangle^{(2)} & = \bigl( \alpha^2_{-\frac{1}{2}} \psi^1_0 - \alpha^1_{-\frac{1}{2}} \psi^2_0 \bigr) |0\rangle^{(2)} \ ,
\label{halfdes1}
\end{align}
and similarly at higher excitation level.

\bibliographystyle{JHEP}

\end{document}